\documentclass[a4paper,fleqn,11pt]{cas-sc}
\usepackage{longtable}
\usepackage{enumitem}
\usepackage{appendix}
\usepackage{multicol}
\usepackage[authoryear]{natbib}
\usepackage{color}
\usepackage{amsmath,amssymb,amsfonts}
\usepackage[linesnumbered,ruled]{algorithm2e}
\usepackage{bm}
\usepackage{subcaption}
\usepackage{booktabs}
\usepackage{cuted}
\usepackage{epstopdf}
\usepackage{epsfig}
\usepackage{subcaption}
\usepackage[english]{babel}
\def\tsc#1{\csdef{#1}{\textsc{\lowercase{#1}}\xspace}}
\tsc{WGM}
\tsc{QE}
\tsc{EP}
\tsc{PMS}
\tsc{BEC}
\tsc{DE}

\newtheorem{definition}{Definition}
\newtheorem{theorem}{Theorem}

\newtheorem{lemma}[theorem]{Lemma}
\newtheorem{assumption}[theorem]{Assumption}
\newtheorem{remark}{Remark}

\begin{document}
\let\WriteBookmarks\relax
\def\floatpagepagefraction{1}
\def\textpagefraction{.001}

\shorttitle{A Parameter Privacy-Preserving Strategy for Mixed-Autonomy Platoon Control}

\shortauthors{Zhou and Yang}

\title [mode = title]{\textbf{A Parameter Privacy-Preserving Strategy for Mixed-Autonomy Platoon Control}}               

\author[1]{Jingyuan Zhou}[orcid=0000-0002-1201-3189]

\ead{jingyuanzhou@u.nus.edu}



\author[1]{Kaidi Yang}[orcid=0000-0001-5120-2866]
\ead{kaidi.yang@nus.edu.sg}

\cormark[1]
\affiliation[1]{organization={Department of Civil and Environmental Engineering, National University of Singapore},
    addressline={1 Engineering Drive 2}, 
    city={Singapore},
    postcode={117576}, 
    country={Singapore}}

\cortext[cor1]{Corresponding author}



\begin{abstract}
It has been demonstrated that leading cruise control (LCC) can improve the operation of mixed-autonomy platoons by allowing connected and automated vehicles (CAVs) to make longitudinal control decisions based on the information provided by surrounding vehicles. However, LCC generally requires surrounding human-driven vehicles (HDVs) to share their real-time states,  which can be used by adversaries to infer drivers' car-following behavior, potentially leading to financial losses or safety concerns. This paper aims to address such privacy concerns and protect the behavioral characteristics of HDVs by devising a parameter privacy-preserving approach for mixed-autonomy platoon control. 
First,  we integrate a parameter privacy filter into LCC to protect sensitive car-following parameters. The privacy filter allows each vehicle to generate seemingly realistic pseudo states by distorting the true parameters to pseudo parameters, which can protect drivers' privacy in behavioral parameters without significantly influencing the control performance. 
Second, to enhance the reliability and practicality of the privacy filter within LCC, we first introduce an individual-level parameter privacy preservation constraint to the privacy filter, focusing on the privacy level of each individual parameter pair. Subsequently, we extend the current approach to accommodate continuous parameter spaces through a neural network estimator.
Third, analysis of head-to-tail string stability reveals the potential impact of privacy filters in degrading mixed traffic flow performance. Simulation shows that this approach can effectively trade off privacy and control performance in LCC. We further demonstrate the benefit of such an approach in networked systems, i.e.,  by applying the privacy filter to a preceding vehicle, one can also achieve a certain level of privacy for the following vehicle. 

\end{abstract}

\maketitle

\section{Introduction}

The transformative technology of connected and automated vehicles (CAVs) presents enormous potential in improving transportation systems~\citep{yang2016isolated,li2017dynamical,feng2018spatiotemporal,tilg2018evaluating,guo2019urban,zhou2021analytical,woo2021flow,tan2022joint,deng2023cooperative,apostolakis2023energy,tan2024connected}. 
Significant research attention has been paid to the longitudinal control of CAVs within a platoon, thanks to the benefits of improving traffic capacity and stability. 
Early studies of CAV longitudinal control focus on cooperative adaptive cruise control (CACC)~\citep{zhang2022memory,dey2015review,xiao2018unravelling,zhang2020cooperative}, which coordinates a platoon of CAVs following a designed head vehicle. 
However, a long transition is expected to be needed to achieve full autonomy, and hence, CAVs and human-driven vehicles (HDVs) will coexist in the near future. 

The key challenge associated with the control of mixed-autonomy platoons lies in the coordination of CAVs, considering the uncertain behavior of HDVs. \citet{orosz2016connected} and \citet{shen2023energy} formulated and proposed control strategies for connected cruise control, where CAVs can utilize the real-time information of the HDVs ahead of them. Later, \citet{wang2021leading} introduced the notion of leading cruise control (LCC), which extends the classical cruise control by allowing CAVs to exploit the information of both leading and following HDVs. In LCC, a central unit (e.g., roadside unit or cloud) collects information from all vehicles in the platoon via vehicle-to-infrastructure (V2I) communications and calculates the optimal control input for CAVs. Various controllers have been proposed to implement LCC, including linear feedback controllers~\citep{wang2021leading}, optimization-based controllers~\citep{chen2021mixed}, data-driven controllers~\citep{wang2022data,wang2022distributed} and deep reinforcement learning-based controllers~\citep{wang2023adaptive,zhou2024enhancing}. These controllers have been demonstrated to be effective in improving string stability and reducing energy consumption while ensuring safety~\citep{zhou2024enhancing}. 

However, LCC requires both HDVs and CAVs to share real-time system states (e.g., positions and velocity) with the central unit, which can consequently store a long history of HDVs' shared states. Such data-sharing can impose privacy risks on these vehicles. The honest but curious central unit, or any adversaries able to access the information stored on it (e.g., by wiretapping the communication channel or attacking the database of the central unit), can exploit the received vehicle states to infer sensitive parameters characterizing HDVs' driving behavior in various traffic scenarios. The leakage of information on HDV driving patterns can lead to economic losses or compromise safety for drivers. For example, \citet{he2018profiling} employed vehicle trajectory data to model driver profiles and developed a personalized insurance pricing model, whereby drivers exhibiting aggressive car-following behavior may face higher insurance rates. Therefore, protecting parameter privacy is crucial to concealing driver behavior and thus preventing potential economic penalties. Moreover, \citet{arbabzadeh2017data} modeled driver safety risk using individual driving data, which shows that the misuse of driver behavior data could compromise the safety of HDVs. Specifically, knowing a driver's typical behavior, attackers could mislead both the human driver and the Advanced Driver Assistance Systems (ADAS), potentially causing safety-critical events \citep{hanselmann2022king}. Due to such concerns, HDVs may not be willing to participate in reporting their states, which may reduce the amount of information the central unit can collect and hence undermine the benefits of LCC. 


To the best of our knowledge, there has been no research on protecting sensitive system parameters of HDVs, especially in the context of mixed-autonomy platoon control. 
The existing literature considering vehicle privacy focuses on location privacy~\citep{sun2013privacy,ma2019real,li2019design,cai2021trajectory,qin2022toward,pan2022privacy,zhang2023privacy,sun2023synthesizing,tsao2022private,tsao2022trust,tsao2023differentially,tan2024privacy,wang2024privacy}, i.e., protecting the location or location sequence of a vehicle from being accessed by adversaries. 
For example, \citet{li2019design} presented a Paillier cryptosystem-based platoon recommendation scheme for head vehicle selection, which effectively ranks head vehicles based on user feedback in a privacy-preserving manner. 
\citet{pan2022privacy} introduced a fault-tolerant encryption-decryption control strategy designed for each vehicle within a platoon, which can effectively preserve the privacy of the location and velocity of individual vehicles while simultaneously ensuring controller stability.
\citet{zhang2023privacy} designed an affine masking-based privacy protection method in the LCC context to protect information from CAVs by transforming their true states and control signals via a low-dimensional affine transformation. However, the method proposed in this work can only be applied to data-enabled predictive control (DeePC) and may not be applicable to other control methods. 
Moreover, although these works can preserve the privacy of vehicle locations, they do not necessarily prevent key system parameters from being inferred due to the law of large numbers~\citep{nekouei2022model}. For example, adversaries can apply classical model identification methods~\citep{teunissen1990nonlinear,wills2008gradient} such as Bayesian filters to estimate the sensitive parameters with high confidence. Although \citet{he2020optimal} attempted to protect the privacy of sensitive parameters of transportation network companies when they share trajectory data, their work considered only the parameters associated with a static optimization problem instead of the parameters of dynamical systems, which can be more challenging to protect due to the temporal dependence of the shared data. 

To address the aforementioned gaps, in this paper, we integrate a statistical parameter privacy filter into the control of mixed-autonomy platoons to distort the information shared by HDVs to protect their car-following parameters. Our proposed method extends \citet{nekouei2022model}, one of the few existing methodological tools to protect the privacy of sensitive parameters of dynamical systems, which generates seemingly realistic synthetic system states by distorting sensitive parameters. 
Although the privacy filter proposed in \citet{nekouei2022model} has been shown to be effective in various applications, such as protecting control gains in the Active Steering Control System~\citep{nekouei2021randomized}, it suffers from three limitations. 
First, the privacy filter was developed for scenarios with small discrete parameter space and can be computationally inefficient if the parameter space is large or continuous, e.g., the car-following parameters in the case of mixed-autonomy platoon control. Second, the privacy guarantee provided by \citet{nekouei2022model} is for the entire distribution of the parameters rather than each possible value. In other words, the resulting privacy protection may be weak for drivers with specific parameters, especially if the parameters appear with a low probability, leading to fairness issues among drivers using this privacy filter. Third, previous applications of the parameter privacy filter have not considered the trade-off between privacy and control performance in networked systems such as vehicle platoons, whereby the perturbations added to the states of one vehicle could influence the string stability of the platoon. To address the first two limitations, we extend \citet{nekouei2022model} by devising a learning-based privacy filter with an individual-level privacy preservation constraint, which can not only accelerate the computation but also provide the privacy guarantee for individual car-following parameters. To address the third limitation, we investigate the trade-offs associated with fuel consumption, velocity error, and head-to-tail string stability \citep{feng2019string} in the mixed-autonomy platoon. The head-to-tail string stability analysis is conducted by constructing the head-to-tail transfer function under privacy filters' perturbations and string stable regions over different pseudo parameters. 

\emph{Statement of Contribution}. The contribution of this paper is three-fold. First, we propose a parameter privacy-preserving strategy to protect the car-following parameters of HDVs from being inferred from their shared data in mixed-autonomy platoon control. This is one of the pioneering works on protecting the privacy of sensitive parameters associated with the dynamic operation of traffic systems. Unlike existing works that merely conceal or perturb vehicles' trajectory data, our approach can protect against inference attacks on sensitive parameters of vehicles' longitudinal driving behavior in mixed-autonomy platoons. Second, we propose a learning-based parameter privacy filter with an individual-level privacy preservation constraint to overcome the inability of the privacy filter to handle continuous parameter space and the potentially weak privacy-preserving performance when applied to specific sensitive parameters. This learning-based estimator efficiently processes continuous parameter input, while the individual-level constraint effectively bounds the entropy of its output, providing an effective solution for preserving privacy in HDVs' car-following model.
Third, we systematically analyze the impact of the proposed parameter privacy filters on control performance, considering the trade-off between privacy and control performance, including fuel consumption, velocity errors, and head-to-tail string stability of the mixed-autonomy platoon. We show that the proposed parameter privacy filters can protect privacy with an acceptable impact on these performance indicators. 

The rest of this paper is organized as follows. Section~\ref{sec: Problem Statement} introduces the problem statement, including the modeling and control of mixed-autonomy platoons and privacy concerns associated with such systems. Section~\ref{sec: Privacy Filter Design} presents the privacy filter designed for mixed traffic to protect the car-following parameters. Section~\ref{sec: Simulation Results} analyses the privacy-utility trade-off. Section~\ref{sec: string stability} analyses the head-to-tail string stability resulting from the parameter privacy filter. Section~\ref{sec: Conclusions and Future Studies} concludes the paper and proposes future research directions.

\section{Problem Statement}
\label{sec: Problem Statement}
\begin{figure}[t]
\centering
\includegraphics[width=16cm]{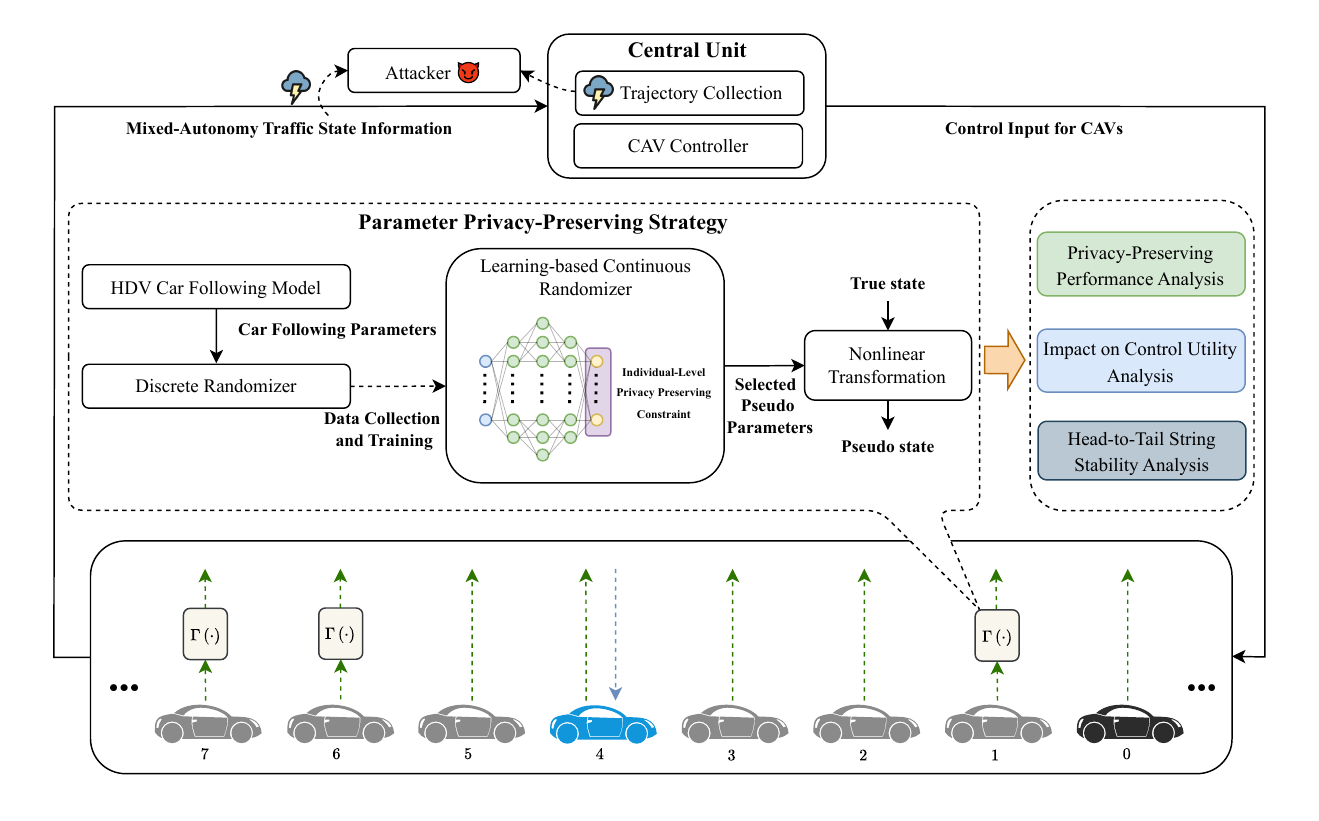}
\caption{Overview of the proposed parameter privacy filter for the mixed-autonomy platoon, where blue vehicles represent CAVs, gray vehicles represent HDVs, and the black vehicle denotes the head vehicle. Some HDVs are equipped with a privacy filter $\Gamma(\cdot)$ to protect the privacy of sensitive car-following parameters. The proposed filter enables HDVs to distort their reported states to pseudo states, thereby protecting sensitive parameters. The trade-offs between privacy and control performance are analyzed. }
\label{overview}
\end{figure}

In this section, we present our problem statement and introduce the privacy concerns associated with mixed-autonomy platoon control. Section~\ref{subsec: Operation of Mixed-Autonomy Platoons} introduces the system modeling for the human drivers and the controller design for mixed-autonomy platoons. Section~\ref{subsec: Privacy Concerns} presents privacy concerns in mixed-autonomy platoons.

\subsection{Operation of Mixed-Autonomy Platoons}
\label{subsec: Operation of Mixed-Autonomy Platoons}
As illustrated in Fig.~\ref{overview}, we consider a mixed-autonomy platoon comprising HDVs and CAVs, whereby the longitudinal actions of HDVs are characterized by the output of car-following models, and the actions of CAVs are determined by a designed controller that relies on the transmitted state information from surrounding HDVs and other CAVs. 

Mathematically, we consider a mixed-autonomy platoon of vehicles indexed by $i\in\Omega$ where $\Omega$ is an ordered set. Let $\Omega_{\mathcal{C}}$ and $\Omega_{\mathcal{H}}$, respectively, denote the sets of CAVs and HDVs within a mixed-autonomy platoon such that $\Omega = \Omega_C\cup\Omega_H$ and $\Omega_C\cap\Omega_H = \emptyset$. Here, $n=|\Omega_{\mathcal{C}}\cup \Omega_{\mathcal{H}}|$ and $m=|\Omega_{\mathcal{C}}|$ denote the number of vehicles and the number of CAVs, respectively. For each vehicle $i \in \Omega$, we characterize its vehicle dynamics by tracking its real-time state $x_i(t) = [s_i(t), v_i(t)]^\top$, where $v_i(t)$ denotes its real-time velocity, and $s_i(t)$ denotes the real-time spacing between vehicle $i$ and its preceding vehicle $i-1$. 

\subsubsection{Modeling of Human Drivers}

The driving behavior for each HDV is governed by a car-following model, such as the Optimal Velocity Model (OVM) \citep{bando1998analysis}, the Intelligent Driver Model (IDM) \citep{treiber2000congested},  the Gipps Model \citep{gipps1981behavioural}, etc. Generally, a car-following model is defined as follows:
\begin{definition}[Car-following Model]
A car-following model is a dynamic model that describes how a following vehicle adjusts its velocity in response to its preceding vehicle. A typical car-following model relates the following vehicle’s acceleration $a_i(t)$ to its current velocity $v_i(t)$, the velocity of the preceding vehicle $v_{i-1}(t)$, and the spacing between the following and preceding vehicles $s_i(t)$, mathematically represented as:
\begin{equation}
    a_i(t) = \mathbf{F} _{\mathbb{\kappa}_i}\left(s_i(t), v_i(t), v_{i-1}(t)\right)
\end{equation}
where the function $\mathbf{F} _{\mathbb{\kappa}_i} (\cdot)$ represents the car-following model of HDV $i$ parametrized by a vector of sensitive car-following parameters $\mathbb{\kappa}_i= \left[\omega_{i,1}, \omega_{i,2}, \ldots, \omega_{i,r}\right]$ representing the driving style, with $r \in \mathbb{N}^+$ indicating the number of sensitive parameters. 
\end{definition}

The continuous parameters $\mathbb{\kappa}_i$ in car-following models typically represent the driving style in a specific traffic scenario. For instance, for a simple linear car-following model $\mathbf{F} _{\mathbb{\kappa}_i}\left(s_i(t), v_i(t), v_{i-1}(t)\right)=\omega_{i,1}\left(s_i(t)-s^\star\right)+\omega_{i,2}\left(v_i(t)-v_{i-1}(t)\right)$, where $s^\star$ represents the equilibrium spacing and $v^\star$ the equilibrium velocity. The sensitive parameters are $\omega_{i,1}$ and $\omega_{i,2}$ that represent the sensitivity to spacing tracking errors and the sensitivity to velocity differences, respectively. A higher sensitivity means that HDVs are more responsive to changes in spacing tracking errors or velocity differences but could potentially lead to system instability if the value is too high.

We then make the following two assumptions regarding the sensitive car-following parameters of HDVs. 
\begin{assumption}[Car-following parameters]\label{asm:cf_para}
    We assume that each HDV knows its own car-following parameters in each traffic scenario (e.g., highway, urban materials, residential areas, etc.).
\end{assumption}

Assumption~\ref{asm:cf_para} is realistic as an HDV can learn its own parameters from its historical states $x_i(t)$ via any system identification methods. Note that we allow such parameters for each HDV to vary across traffic scenarios (e.g., highway, urban materials, residential areas, etc.). Moreover, it is worth noting that although we focus on protecting HDV parameters, the proposed method can be applied in a similar manner to protect parameters associated with CAVs operations (e.g., control gains), which can be seen as a trade secret of the car manufacturer. 

Furthermore, given a population of drivers, their car-following parameters in any given scenario can be characterized by a continuous probability distribution. We make the following assumption about such a distribution.  
\begin{assumption}[Distribution of car-following parameters] \label{asm:para_dist}
We assume the car-following parameter $\kappa$ follows an independent and identical continuous probability distribution over a set $\mathcal{K}$ with a probability density function $\operatorname{P}_{\bm{\kappa}}(\cdot)$, which is publicly known. 
\end{assumption}

Assumption~\ref{asm:para_dist} is realistic since the car-following parameter of each individual driver is her own characteristics and hence independent of other drivers. Furthermore, we assume the distribution of such parameters to be publicly known for two reasons. First, one can always learn such a prior distribution from public naturalistic vehicle trajectory datasets such as the NGSIM dataset \citep{punzo2011assessment}, pNEUMA dataset \citep{barmpounakis2020new}, and Waymo dataset \citep{ettinger2021large,chen2023womd}. Second, this is a conservative assumption because the knowledge of such a distribution can make it easier for attackers to learn the drivers' car-following parameters, and therefore, this assumption enables us to design a stronger privacy-preserving strategy.

\subsubsection{Control of Mixed-Autonomy Platoons}
For the mixed-autonomy platoon containing HDVs $i \in \Omega_\mathcal{\mathcal{H}}$ and CAVs $i \in \Omega_{\mathcal{C}}$, the system dynamics can be written as:
\begin{subequations}
    \begin{align}
    \dot{s}_{i}(t) &=v_{i-1}(t)-v_{i}(t),\\
    \dot{v}_{i}(t) &= 
    \left\{
    \begin{array}{llll}
         &u_i(t), &\quad i \in \Omega_\mathcal{C},\\
         &\mathbf{F} _{\mathbb{\kappa}_i}\left(s_i(t), v_i(t), v_{i-1}(t)\right), &\quad i \in \Omega_\mathcal{H},
    \end{array}
    \right.    
\end{align}\label{eq:system dynamics}
\end{subequations}
where the acceleration rate of CAV $i$ is determined by the control action $u_i(t)$, and the acceleration rate of HDV $i$ is governed by the car-following model $\mathbf{F} _{\mathbb{\kappa}_i}$.

We can rewrite Eq. \eqref{eq:system dynamics} for all the vehicles in a platoon $\Omega = \Omega_{\mathcal{H}}\cup \Omega_{\mathcal{C}}$ as a control affine system:
\begin{equation}
    \dot{x}(t)=f(x(t),v_0(t))+B u(t),
    \label{continuous system}
\end{equation}
where $x(t) = \{x_i(t)\}_{i \in \Omega_{\mathcal{C}}\cup \Omega_{\mathcal{H}}} \in \mathbb{R}_+^{2n}$ and $u(t) = \{u_i(t)\}_{i \in \Omega_{\mathcal{C}}}\in \mathbb{R}_+^m$ represents the states of all vehicles in the platoon and control inputs of all CAVs, respectively. $v_0(t)$ is the velocity of the head vehicle, and $f\left(\cdot\right)$ represents the nonlinear dynamics of HDVs and the linear dynamics of CAVs. System matrix $B$ for CAVs' control input is given as $
    B=\left[e_{2 n}^{2 i_1}, e_{2 n}^{2 i_2}, \ldots, e_{2 n}^{2 i_m}\right] \in \mathbb{R}^{2 n \times m}$, 
where the vector $e^i_{2n}\in\mathbb{R}_+^{2n}$ associated with CAV $i\in \Omega_{\mathcal{C}}$ is a vector with the $2i$-th entry being one and the others being zeros.

In the LCC context, the central unit determines the control action $u_i(t)$ for each CAV $i\in\Omega_{\mathcal{C}}$ based on the error state information transmitted from all vehicles within the platoon. Here, the error state of vehicle $j\in \Omega_{\mathcal{C}}\cup\Omega_{\mathcal{H}}$ is defined as $x^{\text{err}}_j=\left[s_j-s_j^\star,v_j-v^\star\right]$, where $x_j^\star = [s_j^\star,v^\star]$ represents the equilibrium state of the platoon, with $v^\star$ being specified by the head vehicle and $s_j^\star$ calculated by vehicle $j$  by solving $\mathbf{F} _{\mathbb{\kappa}_j}\left(s_j^\star, v^\star, v^\star\right)=0$ from its car-following model.  We represent the error state information received by the CAV from all surrounding vehicles as $y(t)=\left\{\{\bar{x}^{\text{err}}_j(t)\}_{j \in \Omega_\mathcal{H}},\{x^{\text{err}}_j(t)\}_{j \in \Omega_\mathcal{C}}\right\}$. 
Notice that for HDV $j$, here we distinguish the true error state $x^{\text{err}}_j(t)$ and the transmitted error state $\bar{x}^{\text{err}}_j(t)$ because we allow HDVs to perturb their state information from $x_j(t)$ to $\bar{x}_j(t)$ before sharing it with others to protect the privacy of their driving behavior. The perturbed state is determined by a parameter privacy filter, as presented in Section~\ref{sec: Privacy Filter Design}.

In our study, we employ two distinct types of controllers for CAVs, i.e., a centralized data-driven controller and a distributed linear feedback controller, which represent examples of a wide spectrum of platoon controllers. 
The two controllers are described as follows. 

\begin{itemize}
    \item \emph{The centralized data-driven controller:} The controller is based on data-enabled predictive control (DeePC) \citep{coulson2019data}, which is applied to mixed-autonomy platoons with formulations in Appendix~\ref{appendix: DeePC Formulation for Mixed-Autonomy Platoons} \citep{wang2023deep}. As DeePC works in the discrete-time linear system, we thus discretize and linearize the original system dynamics in Eq.~\eqref{continuous system} into a discrete-time linear time-invariant (LTI) system with time step denoted by $t_{\text{s}}$, for which the details are given in Appendix~\ref{appendix: Linearized Discrete-Time System Dynamics for Mixed-Autonomy Traffic}. 
    \item \emph{The distributed linear controller:} Regarding the distributed linear controller, we implement a strategy similar to that described in~\citet{wang2021leading} for each CAV. This strategy leverages the transmitted error states to stabilize the mixed-autonomy platoon, with an aim to maintain an equilibrium state $\left(s_i^\star, v^\star\right)$:
\begin{equation}
\begin{aligned}
u_i(t)=\sum_{j_{\text{CAV}}\in \Omega_\mathcal{C}}\left(\mu_{i,j_{\text{CAV}}} s^{\text{err}}_{j_{\text{CAV}}}(t)+\eta_{i,j_{\text{CAV}}} v^{\text{err}}_{j_{\text{CAV}}}(t)\right)+\sum_{j_{\text{HDV}} \in \Omega_\mathcal{H}}\left(\mu_{i,j_{\text{HDV}}} \bar{s}^{\text{err}}_{j_{\text{HDV}}}(t)+\eta_{i,j_{\text{HDV}}} \bar{v}^{\text{err}}_{j_{\text{HDV}}}(t)\right),
\end{aligned}
\label{linear controller}
\end{equation}
where $\mu_{i,j_{\text{CAV}}},\eta_{i,j_{\text{CAV}}}$ are the feedback gains corresponding to the states of CAVs $j_{\text{CAV}}\in\Omega_{\mathcal{C}}$, and $\mu_{i,j_{\text{HDV}}},\eta_{i,j_{\text{HDV}}}$ are the feedback gains corresponding to the surrounding HDVs $j_{\text{HDV}} \in \Omega_\mathcal{H}$, which are chosen based on the string stability conditions following~\citet{wang2021leading}. 
\end{itemize}

\subsection{Privacy Concerns Associated With Mixed-Autonomy Platoons}
\label{subsec: Privacy Concerns}
From the system operations described in Section~\ref{subsec: Operation of Mixed-Autonomy Platoons}, we can see that the major privacy concern lies in the sharing of HDVs' state information with the central unit, which could potentially be used by adversaries to infer HDV driving styles (characterized by sensitive car-following parameters). In the context of mixed-autonomy platoons, we prioritize parameter privacy over location privacy for three main reasons. First, car-following parameters in HDVs reveal specific driving styles, and the leakage of such information may allow adversaries to manipulate insurance pricing or perform cyber and physical attacks, causing potential economic losses and safety concerns for drivers.
Second, location privacy is more important at the trip level, where routing and origin-destination information can reveal sensitive information about travelers. However, in our context, where the controller knows the roads on which the vehicles travel, precise GPS locations would not provide additional sensitive information about travelers other than the driving style characterized by sensitive parameters. 
Third, protecting travelers' exact GPS locations is not sufficient to preserve the privacy of sensitive parameters due to the law of large numbers \citep{nekouei2022model}, as adversaries can always apply classical model identification methods \citep{teunissen1990nonlinear} to estimate the sensitive parameters with high confidence. 

Next, we formalize our assumptions about the adversaries that aim to infer the sensitive car-following parameters.

\begin{assumption}[Adversaries]\label{asm:adv}
    Adversaries include any external attackers who can access the data transmitted from an HDV to the central unit (e.g., by wiretapping the communication channel or attacking the database of the central unit). 
    They aim to estimate the car-following parameters of an HDV using the following information: (1) transmitted states of all vehicles within the platoon, represented by $y(t) = \left\{ \bar{x}^{\text{err}}_i \right\}_{i\in \Omega_{\mathcal{H}}} \cup \left\{ x^{\text{err}}_j \right\}_{j\in \Omega_{\mathcal{C}}}$, and (2) the prior distribution of car-following model parameters characterized by $\operatorname{P}_{\kappa}(\cdot)$.  Specifically, adversaries can leverage any parameter estimation model to learn car-following parameters  $\kappa_i=\left[\omega_{i1}, \omega_{i2}, \ldots, \omega_{ir}\right]$ for HDV $i\in\Omega_{\mathcal{H}}$ that minimizes the estimation performance metric of normalized root mean square error (RMSE), i.e., $\mathrm{E}\left[\sqrt{\sum_{l=1}^r\left(\frac{\hat{\omega}_{il}-\omega_{il}}{\omega_{il}}\right)^2}\right]$. 
\end{assumption}

We make the following remarks regarding Assumption~\ref{asm:adv}. First, as presented in the previous section, the central unit can access and store the transmitted error states from each HDV and then send the control command to each CAV. Hence, external attackers can attack the central units' communication channels or databases to illegally access these transmitted states and use them to infer the true parameters of each HDV via a parameter estimator.
Second, certain CAVs are admittedly equipped with sensors capable of capturing the real states of surrounding vehicles, which may be potentially employed to infer sensitive parameters of surrounding vehicles. However, in practice, an HDV is typically in the sensing range of the CAV for a relatively short period of time, and therefore, the data collected about the HDV involves only short-term trajectories for particular scenarios, which can be insufficient for accurate and holistic estimation of HDV behavior. Hence, we argue that the primary privacy concerns of HDV drivers about such a system lie in the sharing of their long-term and comprehensive trajectory data to the central unit rather than the unavoidable detection of surrounding CAVs. 
Third, we use normalized RMSE as the attacker's metric, which is simply the RMSE divided by the variance of the true data. It is preferred over regular RMSE because it provides a more standardized way to assess the model's performance, especially when dealing with data from different scales and magnitudes. This choice aligns well with our scenario, where $\kappa$ encompasses multiple sensitive parameters with different scales, making normalized RMSE a more suitable evaluation metric. 

Next, we devise a parameter privacy filter to protect the sensitive parameters associated with HDV driving behavior. 

\section{Parameter Privacy Filter for Mixed-Autonomy Platoons}
\label{sec: Privacy Filter Design}
This section introduces the parameter privacy filter for LCC to protect the privacy of sensitive car-following parameters of HDVs, i.e., the parameters $\kappa_i$, against adversaries. Section~\ref{subsec: General Framework} describes the methodological framework of the parameter privacy filter. Section~\ref{subsec: Randomizer} presents the details of the randomizer, which is responsible for generating pseudo parameters based on the true sensitive parameters. Section~\ref{subsec: Nonlinear Transformation} introduces the nonlinear transformation employed to utilize these pseudo parameters in synthesizing pseudo states.

\subsection{General Framework}
\label{subsec: General Framework}
With the assumptions given in Section~\ref{subsec: Privacy Concerns}, we next present the parameter privacy filter as in Fig.~\ref{overview}, a randomized mechanism denoted by $\Gamma_i\left(\cdot\right)$ that protects sensitive parameters $\kappa_i$ by generating pseudo-states $\bar{x}_i$ from (i) historical trajectory of true states $X_i=\{x_i(t_s)\}_{t_s=1}^T$, (ii) true car-following parameters $\kappa_i$, and (iii) external inputs (i.e., the velocity of the preceding vehicle $v_{i-1}$). Specifically, the privacy filter for vehicle $i \in \Omega_\mathcal{H}$ can be written as 
\begin{equation}
    \bar{x}_i = \Gamma_i\left(X_i,v_{i-1},\kappa_i\right), i \in \Omega_\mathcal{H}.
    \label{eq: states after privacy filter}
\end{equation}

Let variable $\bar{X}_i=\{\bar{x}_i(t_s)\}_{t_s=1}^T$ represents a sequence of pseudo states for HDV $i$. Here, we use the bolded variables $\bm{\kappa}$ and $\bm{\bar{X}}_i$ to represent the sensitive parameters and pseudo states as random variables, as described in Assumption~\ref{asm:para_dist} and Eq.~(\ref{eq: states after privacy filter}), and $\kappa$ and $\bar{X}_i$ to represent their specific realizations.    
We find the privacy filter Eq.~\eqref{eq: states after privacy filter} by solving the following optimization problem:
\begin{subequations}
\begin{align}
\min_{\Gamma_i}\quad & \mathbb{E}_{\bm{\bar{X}}_i}\text{D}\left(X_i,\bm{\bar{X}}_i\right) \label{eq: general objective} \\
\text{s.t.}\quad& \mathrm{I}\left[\bm{\kappa} ; \bm{\bar{X}}_i\right]  \leq I_0, \label{eq: general MI}
\end{align} \label{eq:general_opt}
\end{subequations}
where the 
objective function~Eq.\,\eqref{eq: general objective} minimizes the expected discrepancy between the true states and pseudo states. Constraint~Eq.\,\eqref{eq: general MI} sets an upper bound $I_0$ on the mutual information $\mathrm{I}[\cdot]$ between true parameter $\bm{\kappa}$ and pseudo states $\bar{X}_i$. The mutual information quantifies the amount of information about $\bm{\kappa}$ that can be inferred from $\bm{\bar{X}}_i$. A large value of $\mathrm{I}[\cdot]$ results in a higher risk of information leakage of true sensitive parameters, while a small value of $\mathrm{I}[\cdot]$ can ensure a higher privacy level. Here, for continuous values of $\kappa$ and $\bar{X}_i$, the mutual information is represented as follows ($\int$ can be changed to $\sum$ for discrete variables): 
\begin{align}
    \mathrm{I}[\boldsymbol{\kappa} ; \bm{\bar{X}}_i]=& \int_{\bm{\kappa},\bm{\bar{X}}_i} \operatorname{P}_{\boldsymbol{\bar{X}_i,\kappa}}\left(\bar{X}_i,\kappa\right)
 \log \frac{\operatorname{P}_{\boldsymbol{\bar{X}_i,\kappa}}
 \left(\bar{X}_i,\kappa\right)}{\operatorname{P}_{\boldsymbol{\bar{X}_i}}\left(\bar{X}_i\right) \operatorname{P}_{\boldsymbol{\kappa}}\left(\kappa\right)}\mathrm{d}\kappa\mathrm{d}\bar{X}_i.
\end{align}
where $\operatorname{P}_{\boldsymbol{x}}(x)$ represents the probability density function of the corresponding random variable $\boldsymbol{x}$.

 Overall, the optimization problem Eq.\,\eqref{eq:general_opt} allows for the trade-off between privacy (represented by Eq.\,\eqref{eq: general objective}) and the accuracy of reported states (represented by Eq.\,\eqref{eq: general MI}). Specifically, one (the driver or vehicle manufacturer) can always choose an appropriate value of $I_0$ (the upper bound of privacy leakage) to ensure that the accuracy of reported states is within an acceptable range. In fact, as we demonstrate in the case study in Section \ref{subsubsec: Distribution of the Discrepancy between Pseudo States and True States}, the discrepancy between reported states and true states is relatively small. This makes our proposed privacy filter significantly different from cyber attacks, which aim to disturb the systems (e.g., causing safety issues). 

 Moreover, we present Remark \ref{rmk:safety} to justify that our proposed privacy filter, despite the distortion of HDV states, will not compromise the safety of HDVs or CAVs. 

 \begin{remark}[Safety considerations]\label{rmk:safety}
The proposed privacy filter does not compromise the safety of HDVs or CAVs for two reasons. First, since each CAV is equipped with onboard sensors to detect the velocity of the preceding vehicle and the spacing to the preceding vehicle, it can accurately represent the safety constraints in its decision-making process. 
On the other hand, HDVs' decisions are determined by drivers without using any transmitted states from other vehicles, so they are not affected by the privacy filter. 
Second, the level of distortion introduced by the privacy filter is small and adjustable since the privacy filter aims to minimize the discrepancy between the pseudo states and true states while ensuring that the mutual information is below a user-selected upper bound. 
To support this claim, we conduct simulations to assess the discrepancy between the pseudo and true states in Section \ref{subsubsec: Distribution of the Discrepancy between Pseudo States and True States}.
\end{remark}

The optimization problem Eq.\,\eqref{eq: general objective}-Eq.\,\eqref{eq: general MI} is challenging to solve because the calculation of mutual information $\mathrm{I}\left[\bm{\kappa} ; \bm{\bar{X}}_i\right]$ is inherently difficult due to the high dimension of $\bm{\bar{X}}_i$. 
Hence, we follow a heuristic approach proposed in \citet{nekouei2022model} that solves the optimization problem in a two-step manner. As illustrated in Fig. \ref{overview}, the privacy filter consists of two components: (i) a randomizer that distorts the true sensitive parameters $\kappa_i$ to generate pseudo parameters $\bar{\kappa}_i$ according to a probabilistic mapping $\pi(\bar{\kappa}_i|\kappa_i)$, and (ii) a nonlinear transformation that produces pseudo states $\bar{x}_i$ from the pseudo parameters $\bar{\kappa}_i$, true states $x_i$, and the velocity of the preceding vehicle $v_{i-1}$, such that adversaries cannot infer the true parameters from the pseudo error states. 

We next proceed to present the details of the two steps in Section~\ref{subsec: Randomizer} and Section~\ref{subsec: Nonlinear Transformation}, respectively. 

\subsection{Randomizer}
\label{subsec: Randomizer}
The randomizer generates a pseudo parameter $\bar{\kappa} \in \bar{\mathcal{K}}$ using the true parameter $\kappa \in \mathcal{K}$ according to a probabilistic mapping $\pi\left(\bar{\kappa}\mid\kappa \right)$. 
Note that it is more desirable if the set of pseudo states is discrete because this can help reduce computation complexity and better protect parameter privacy. Then, the randomizer is formulated as the following optimization problem in scenarios with continuous true sensitive parameters: 
\begin{subequations}
    \begin{align}
    \min_{\pi(\bar{\kappa}\mid \kappa)} &\int_{\kappa \in \mathcal{K}} \sum_{\bar{\kappa} \in \bar{\mathcal{K}}} \pi(\bar{\kappa}\mid \kappa)\operatorname{P}_{\bm{\kappa}}\left(\kappa\right)e\left(\kappa,\bar{\kappa}\right)\mathrm{d}\kappa \label{eq: continuous optimization}\\
    \text{s.t.} &\sum_{\bar{\kappa}\in\bar{\mathcal{K}}}\pi\left(\bar{\kappa}\mid \kappa\right) =1, \forall \kappa\in\mathcal{K}, \label{eq:con1_c}\\
    &\mathrm{I}[\bar{\boldsymbol{\kappa}};\boldsymbol{\kappa}]\leq I_0, \label{eq:con2_c}\\
    &\pi\left(\bar{\kappa}\mid \kappa\right)\geq 0,\forall \kappa\in\mathcal{K},\bar{\kappa} \in \bar{\mathcal{K}}, \label{eq:con3_c}
    \end{align} \label{eq:randomizer_general}
\end{subequations}
\noindent where the bolded variable $\bm{\kappa}$ and $\bm{\bar{\kappa}}$ represent the true and pseudo parameters as random variables. Constraints~Eq.\,\eqref{eq:con1_c} and~Eq.\,\eqref{eq:con3_c} ensure that the probabilistic mapping $\pi\left(\cdot \mid \cdot \right)$ is a probabilistic distribution. The objective function~Eq.\,\eqref{eq: continuous optimization} minimizes the expected errors in the reported measurement due to the parameter distortion of the privacy filter, where the error function $e(\kappa,\bar{\kappa})$ can be calculated by:
\begin{align}
e(\kappa,\bar{\kappa})=\frac{1}{T} \sum_{t_{\text{s}}=1}^T \mathbb{E}\left.\left[\left\|x_i(t_{\text{s}})-\bar{x}_i(t_{\text{s}})\right\|^2\right|\boldsymbol{\kappa}=\kappa,\boldsymbol{\bar{\kappa}}=\bar{\kappa}\right],
\end{align}
where the variable $\left\{x_i(t_{\text{s}})\right\}_{{t_{\text{s}}}=1}^T$ represents a sampled trajectory (i.e., sequence of states) generated from the true parameters $\boldsymbol{\kappa}$ at discretized time steps with $T$ time intervals indexed by ${t_{\text{s}}}$, and $\left\{\bar{x}_i(t_{\text{s}})\right\}_{{t_{\text{s}}}=1}^T$ represents the sampled trajectory generated from the pseudo parameters $\bar{\boldsymbol{\kappa}}$ at discretized time steps with $T$ time intervals indexed by ${t_{\text{s}}}$, both using Monte Carlo sampling. 

Eq. \eqref{eq:con2_c} follows Eq. \eqref{eq: general MI} to provide an upper bound for the mutual information, which can be given as:
\begin{align}
\mathrm{I}[\boldsymbol{\kappa} ; \bar{\boldsymbol{\kappa}}] & =  \int_{\kappa \in \mathcal{K}}\sum_{\bar{\kappa}\in \bar{\mathcal{K}}} \operatorname{P}_{\bm{\kappa},\bm{\bar{\kappa}}}\left(\kappa,\bar{\kappa}\right)
 \log \frac{\operatorname{P}_{\bm{\kappa},\bm{\bar{\kappa}}}\left(\kappa,\bar{\kappa}\right)}
{\operatorname{P}_{\boldsymbol{\kappa}}\left(\kappa\right) \operatorname{P}_{\boldsymbol{\bar{\kappa}}}\left(\bar{\kappa}\right)} \text{d}\kappa\notag \\
  &= \int_{\kappa \in \mathcal{K}}\sum_{\bar{\kappa} \in \bar{\mathcal{K}}} \pi\left(\bar{\kappa} \mid \kappa\right)\operatorname{P_{\bm{\kappa}}}\left(\kappa\right)\log \frac{\pi\left(\bar{\kappa} \mid \kappa\right)}{\int_{\kappa^\prime \in \mathcal{K}}\pi\left(\bar{\kappa} \mid \kappa^\prime\right)\operatorname{P_{\bm{\kappa}}}\left(\kappa^\prime\right)\mathrm{d}\kappa^\prime}\mathrm{d}\kappa,\label{eq: privacy constraint}
\end{align}
where $\operatorname{P}_{\bm{\kappa}}(\kappa)$ is the probability density function for $\kappa$, and the last equation in Eq.(\ref{eq: privacy constraint}) holds because of $\operatorname{P}_{\bm{\kappa},\bm{\bar{\kappa}}}\left(\kappa,\bar{\kappa}\right) = \pi\left(\bar{\kappa} \mid \kappa\right)\operatorname{P_{\bm{\kappa}}}\left(\kappa\right)$. Since the pseudo states are generated from the pseudo parameters, by the data processing inequality \citep{cover1999elements}, the mutual information between the true parameters $\bm{\kappa}$ and pseudo states $\bm{\bar{X}_i}$ is upper bounded by the mutual information between the true and pseudo parameters, i.e., $\mathrm{I}\left[\boldsymbol{\kappa} ; \boldsymbol{\bar{X}_i}\right] \leq \mathrm{I}[\boldsymbol{\kappa} ; \boldsymbol{\bar{\kappa}}]$.
This relationship, together with the privacy preservation constraint~Eq.\,\eqref{constraint3}, implies $\mathrm{I}\left[\boldsymbol{\kappa} ; \boldsymbol{\bar{X}_i}\right] \leq I_0$, i.e., the output of the privacy filter provides an upper-bounded of information one can infer about the car-following parameters $\kappa$.

However, solving the optimization problem represented in Eq.~\eqref{eq:randomizer_general} proves challenging. We present two approaches to address this issue. The first approach discretizes the true parameter set $\mathcal{K}$ to convert the optimization problem in Eq. \eqref{eq:randomizer_general} into a convex optimization problem \citep{nekouei2022model}, as detailed in Section \ref{subsubsec: Discrete Randomizer}. The second approach employs a neural network to approximate the probabilistic mapping, which is discussed in Section \ref{subsubsec: Extension of Parameter Privacy Filter}.


\subsubsection{Discretized Randomizer with Aggregated-Level Privacy Guarantee}
\label{subsubsec: Discrete Randomizer}
The first approach follows the canonical randomizer proposed by \cite{nekouei2022model}, which discretizes the continuous set of true parameters $\mathcal{K}$ into multiple subsets denoted by $\mathcal{D}$, with the mean value of each subset being selected as its representative value used in the corresponding subset. 
With such discretization, the optimization problem associated with the randomizer in Eq. \eqref{eq:randomizer_general} can be converted into the following optimization problem:
\begin{subequations}
\begin{align}
\min_{\left\{\pi\left(\bar{\kappa} \mid \kappa\right)\right\}_{\bar{\kappa} \in \bar{\mathcal{K}}, \kappa \in \mathcal{D}}} & 
\sum_{\kappa\in \mathcal{D}, \bar{\kappa} \in \bar{\mathcal{K}}} \pi(\bar{\kappa}\mid \kappa)\operatorname{P}_{\bm{\kappa}}\left(\kappa\right)e\left(\kappa,\bar{\kappa}\right)
\label{objective} \\
\text{s.t.}\quad& \pi\left(\bar{\kappa} \mid \kappa \right) \geq 0, \quad \forall \kappa \in \mathcal{D}, \forall \bar{\mathcal{\kappa}} \in \bar{\mathcal{K}}, \label{constraint1}\\
& \sum_{\bar{\kappa} \in \bar{\mathcal{K}}} \pi\left(\bar{\kappa} \mid \kappa \right)=1, \quad \forall \kappa \in \mathcal{D}, \quad  \label{constraint2}\\
& \mathrm{I}[\bm{\kappa} ; \bm{\bar{\kappa}}]  \leq I_0, \label{constraint3}
\end{align} \label{eq:opt-randomizer-discrete}
\end{subequations}
\noindent where the key difference compared to Eq. \eqref{eq:randomizer_general} lies in replacing the integration over a continuous set with the summation over a discrete set.
Constraint~Eq.\,\eqref{constraint3} provides an upper bound $I_0$ for the mutual information between $\boldsymbol{\kappa}$ and $\boldsymbol{\bar{\kappa}}$, i.e.,:
\begin{equation}
\begin{aligned}
\mathrm{I}[\boldsymbol{\kappa} ; \boldsymbol{\bar{\kappa}}]
=\sum_{\kappa \in \mathcal{D}, \bar{\kappa} \in \bar{\mathcal{K}}} \pi\left(\bar{\kappa} \mid \kappa\right)\operatorname{P_{\bm{\kappa}}}\left(\kappa\right)\log \frac{\pi\left(\bar{\kappa} \mid \kappa\right)}{\sum_{\kappa^\prime \in \mathcal{D}}\pi\left(\bar{\kappa} \mid \kappa^\prime\right)\operatorname{P_{\bm{\kappa}}}\left(\kappa^\prime\right)},
\end{aligned}
\label{mutual information}
\end{equation}
which is a convex function of $\pi(\cdot|\cdot)$. According to Theorem 3 in~\citet{nekouei2022model}, with the mutual information upper-bound $I_0$, the error probability of any parameter estimator used by the adversaries can be lower bounded, i.e., 
\begin{equation}
\operatorname{Pr}\left(\boldsymbol{\kappa} \neq \boldsymbol{\hat{\kappa}}\right) \geq \frac{\mathrm{H}[\boldsymbol{\kappa}]-I_0-1}{\log |\mathcal{D}|},
\end{equation}
where $\left|\mathcal{D}\right|$ is the cardinality of $\mathcal{D}$, and $\mathrm{H}[\boldsymbol{\kappa}]$ is the discrete entropy of $\boldsymbol{\kappa}$. 
It is evident that the lower bound of $\operatorname{Pr}\left(\boldsymbol{\kappa} \neq \boldsymbol{\bar{\kappa}}\right)$ increases as $I_0$ decreases, which means a small $I_0$ can reduce the chance that an attacker reliably estimates the car-following parameters. 

The optimization problem Eq.~\eqref{eq:opt-randomizer-discrete} is a convex optimization problem since both the objective and constraints Eq.~\eqref{constraint1}-Eq.~\eqref{constraint2} are linear and the mutual information at the left-hand side of Eq.~\eqref{constraint3} is convex. 

Nevertheless, we identify two limitations associated with the canonical randomizer as proposed in \citet{nekouei2022model}, which may limit the application of the randomizer for mixed-autonomy platoons.
First, such a discretization approach is suitable for cases where the number of variables is not too large (e.g., $\kappa$ has a low dimension). Otherwise, finding the probabilistic mapping can still become intractable due to the extensive computational resources required for Monte Carlo sampling and a significantly large number of decision variables. 
Second, constraint Eq.\,\eqref{constraint3} provides an upper bound for the amount of information leaked by sharing the pseudo states $\bar{X}$, which works in an aggregated manner over all parameters in $\mathcal{K}$ without imposing privacy guarantee for any specific parameter. Therefore, it is likely that privacy is not well guaranteed for certain parameters, especially those that appear in $\mathcal{K}$ with a low probability, leading to fairness issues among drivers. 
 
To address these two limitations, we extend the discretization-based approach to a learning-based approach with individual-level guarantees in Section~\ref{subsubsec: Extension of Parameter Privacy Filter}.

\subsubsection{Learning-Based Continuous Randomizer with Individual-Level Privacy Guarantee}
\label{subsubsec: Extension of Parameter Privacy Filter}
In this section, we introduce a learning-based approach for accelerating the solution process of the randomizer to facilitate the online deployment of the parameter privacy filter while strengthening the privacy guarantee for individual parameters. Specifically, we first strengthen the privacy guarantee using an individual-level privacy-preserving constraint for individual parameter $\kappa \in \mathcal{K}$. Then, we design a learning-based approach to interpolate the probabilistic mapping $\pi(\bar{\kappa}|\kappa) $ over a continuous parameter space $\kappa \in \mathcal{K}$. 
 
\begin{figure}[t]
    \centering
    \includegraphics[width=8cm]{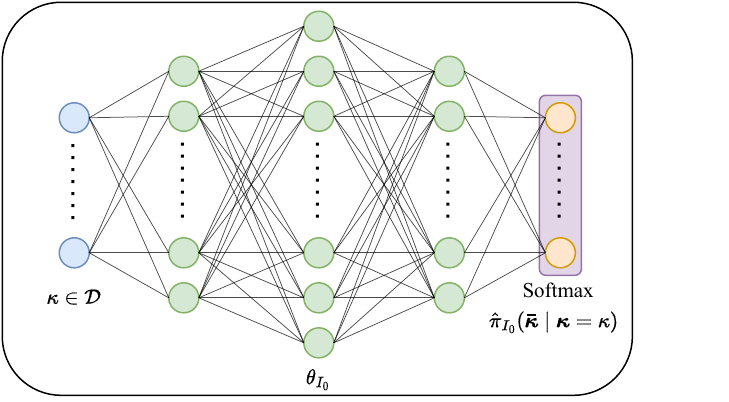}
    \caption{Structure of the neural network-based randomizer.}
    \label{fig: NNR}
\end{figure}

\vspace{0.3em}\noindent \textbf{(1) Individual-Level Privacy Preservation Constraint}.  
By the properties of mutual information and conditional entropy, 
we can rewrite Eq.\,\eqref{constraint3} as: 
\begin{align}    
H(\boldsymbol{\bar{\kappa}}) - I_0 \leq H(\boldsymbol{\bar{\kappa}})  - \mathrm{I}\left[\boldsymbol{\kappa} ; \boldsymbol{\bar{\kappa}}\right] = H(\boldsymbol{\bar{\kappa}}  \mid \boldsymbol{\kappa})=\int_{\kappa \in \mathcal{K}} \operatorname{P}_{\bm{\kappa}}(\kappa) H(\boldsymbol{\bar{\kappa}} \mid \boldsymbol{\kappa}=\kappa)\mathrm{d}\kappa, \label{eq:mi_entropy}
\end{align}
where $H(\boldsymbol{\bar{\kappa}})$ represents the entropy of random variable $\boldsymbol{\bar{\kappa}}$, i.e., $H(\boldsymbol{\bar{\kappa}})=-\sum_{\bar{\kappa}\in\bar{\mathcal{K}}}\operatorname{P}_{\bm{\bar{\kappa}}}(\bar{\kappa})\log \left(\operatorname{P}_{\bm{\bar{\kappa}}}(\bar{\kappa})\right)$, and $H(\boldsymbol{\bar{\kappa}} \mid \boldsymbol{\kappa})$ represents the  entropy of $\boldsymbol{\bar{\kappa}}$ conditional on $\boldsymbol{\kappa}$. The first inequality of Eq.(\ref{eq:mi_entropy}) follows Eq.\,\eqref{constraint3}, the second equality of Eq.(\ref{eq:mi_entropy}) follows the property of mutual information, and the third equality is derived by the definition of conditional entropy. 

Note that Eq.\,\eqref{eq:mi_entropy} is an aggregated-level constraint that aggregates over all parameters $\kappa \in \mathcal{K}$ and hence provides no guarantee for the privacy level of any specific parameter. To address this issue, we strengthen the inequality by requiring that for any specific parameter $\kappa$, the conditional entropy $H(\boldsymbol{\bar{\kappa}}\mid \bm{\kappa}=\kappa)$ satisfies the \emph{individual-level privacy preservation constraint} represented by Eq.~\eqref{eq: Individual-Level privacy preservation constraint}: 
\begin{equation}
H(\boldsymbol{\bar{\kappa}} \mid \boldsymbol{\kappa}=\kappa) = -\sum_{i=1}^n \hat{\pi}_{I_0}\left(\bm{\bar{\kappa}}=\bar{\kappa}^i\mid \bm{\kappa}=\kappa \right) \log \hat{\pi}_{I_0}\left(\bm{\bar{\kappa}}=\bar{\kappa}^i\mid \bm{\kappa}=\kappa \right) \geq H(\boldsymbol{\bar{\kappa}})-I_0.
\label{eq: Individual-Level privacy preservation constraint}
\end{equation}


Then, we can replace the constraint Eq.\,\eqref{constraint3} with Eq.\,\eqref{eq: Individual-Level privacy preservation constraint} to formulate the optimization problem with the individual-level privacy constraint. Notice that the entropy $H(\boldsymbol{\bar{\kappa}})$ on the ride-hand side of Eq.\,\eqref{eq: Individual-Level privacy preservation constraint} replies on the decision variable, i.e., the probabilistic mapping $\pi(\cdot|\cdot)$, which makes this constraint not convex. Nevertheless, since the left-hand side of Eq.\,\eqref{eq: Individual-Level privacy preservation constraint} is concave, and $H(\boldsymbol{\bar{\kappa}})$ is a scalar independent of the true parameter $\kappa$, we can solve this optimization problem by enumerating possible values of $H(\boldsymbol{\bar{\kappa}})$ until the resulting probabilistic mapping $\pi(\cdot|\cdot)$ is consistent with the enumerated value of $H(\boldsymbol{\bar{\kappa}})$. 

Next, we present the learning-based approach to accelerate the solution process of the randomizer with continuous true parameters. 

\vspace{0.3em} \noindent \textbf{(2) Neural Network-Based Randomizer}. As depicted in Fig. \ref{fig: NNR}, to efficiently solve the optimization problem Eq.~\eqref{eq:randomizer_general}, we design a neural network-based estimator $h_{\theta_{I_0}}\left(\kappa \right),~\kappa \in \mathcal{K}$ parameterized by $\theta_{I_0}$ to approximate the probabilistic mapping. Specifically, the input to this neural network is a true parameter $\kappa \in \mathcal{K}$, and the output of the neural network is processed by a softmax function to give a probabilistic distribution $\pi(\cdot|\boldsymbol{\kappa}=\kappa)$, where each element corresponds to a possible pseudo parameter $\bar{\kappa} \in \bar{\mathcal{K}}$. 
Mathematically, let the values of the neural network's output layer be denoted by $m_1, m_2, \cdots, m_n \in \mathbb{R}$, where $n=|\bar{\mathcal{K}}|$ is the cardinality of the set of pseudo parameters. 
Then, the output of the softmax layer for element $i\in\left\{1,...,n\right\}$ is given as:
\begin{equation}
    \hat{\pi}_{I_0}\left(\bm{\bar{\kappa}}=\bar{\kappa}^i\mid \bm{\kappa}=\kappa \right)=\frac{e^\frac{m_i}{T_{\text{scale}}}}{\sum_{i=1}^n e^\frac{m_i}{T_{\text{scale}}}},
    \label{eq: p with t}
\end{equation}
 where $T_{\text{scale}}$ denotes the designed scaling factor, which is set to $1$ by default. 

Such a neural network can be used to estimate the probability distribution of pseudo parameters given true parameters $\kappa \in \mathcal{K}$. 
We then describe the training procedure for such a neural network-based estimator. \vspace{-0.6em}

\begin{itemize}
    \item \emph{Training dataset}. We collect the training dataset using the randomizer with discrete parameters with the individual-level privacy constraint. Specifically, we first discretize the continuous parameter set into multiple subsets, which could follow certain rules (e.g., grids) or stochastic sampling. We select the representative value of each subset as its mean value and solve the optimization problem Eq.\,\eqref{eq:opt-randomizer-discrete} with the individual-level privacy constraint Eq.\,\eqref{eq: Individual-Level privacy preservation constraint} with these representative values denoted by set $\mathcal{D}$ to obtain the corresponding probabilistic mapping $\pi(\cdot)$ with each privacy level $I_0$. Then, the set of $\{(\kappa, \pi(\bm{\bar{\kappa}}|\kappa))\}_{\kappa\in \mathcal{D}}$ is used as the training dataset. We can further repeat the aforementioned procedure several times to randomly generate multiple discretized parameter sets $\mathcal{D}$ and combine the resulting training datasets. \vspace{-0.6em} 

\item \emph{Loss function}. Since our objective is to match the output distribution $\hat{\pi}_{I_0}\left(\bar{\boldsymbol{\kappa}}\mid\boldsymbol{\kappa}=\kappa\right)$ to a given distribution $\pi_{I_0}(\boldsymbol{\bar{\kappa}} \mid \boldsymbol{\kappa}=\kappa)$, we employ the Kullback-Leibler (KL) divergence \citep{van2014renyi} as the loss function. The KL divergence is utilized to measure the distance between two distributions and is given by:
\end{itemize}
\begin{equation}
D_{\mathrm{KL}}\left(\hat{\pi}_{I_0}\left(\bar{\boldsymbol{\kappa}}\mid\boldsymbol{\kappa}=\kappa\right) \| \pi_{I_0}(\boldsymbol{\bar{\kappa}} \mid \boldsymbol{\kappa}=\kappa)\right)=\sum_{\bar{\kappa} \in \bar{\mathcal{K}}}\sum_{\kappa \in \mathcal{D}} \hat{\pi}_{I_0}\left(\bar{\boldsymbol{\kappa}}=\bar{\kappa}\mid\boldsymbol{\kappa}=\kappa\right)\log \left(\frac{\hat{\pi}_{I_0}\left(\bar{\boldsymbol{\kappa}}=\bar{\kappa}\mid\boldsymbol{\kappa}=\kappa\right)}{\pi_{I_0}\left(\bar{\boldsymbol{\kappa}}=\bar{\kappa}\mid\boldsymbol{\kappa}=\kappa\right)}\right).
\end{equation}

However, the constraint Eq.\,\eqref{eq: Individual-Level privacy preservation constraint} is not guaranteed to be satisfied for the trained neural network. To address this issue, we modify the softmax function Eq.\,\eqref{eq: p with t} by locally adjusting the scaling factor $T_{\text{scale}}$. 
Specifically, by integrating Eq.\,\eqref{eq: p with t} into Eq.\,\eqref{eq: Individual-Level privacy preservation constraint}, we can rewrite the individual-level privacy preservation constraint Eq.~\eqref{eq: Individual-Level privacy preservation constraint} as: 
\begin{equation}
    -\sum_{i=1}^n \frac{e^\frac{m_i}{T_{\text{scale}}}}{\sum_{i=1}^n e^\frac{m_i}{T_{\text{scale}}}} \log \frac{e^\frac{m_i}{T_{\text{scale}}}}{\sum_{i=1}^n e^\frac{m_i}{T_{\text{scale}}}} \geq -I_0+H(\boldsymbol{\bar{\kappa}}), \label{eq: Individual-Level privacy constraint}
\end{equation}
We can verify whether the output of the neural network $\{m_i\}_{i=1}^n$ satisfies the individual-level privacy constraint Eq.~\eqref{eq: Individual-Level privacy constraint}. If Eq.\,\eqref{eq: Individual-Level privacy constraint} does not hold, we adjust $T_{\text{scale}}$ such that the equality holds in Eq.\,\eqref{eq: Individual-Level privacy constraint}, which means that the modified probabilistic mapping Eq.\,\eqref{eq: p with t} with scaling factor $T_{\text{scale}}$ satisfies the individual-level privacy preservation constraint. Note that since the individual-level privacy preservation constraint is naturally satisfied in the training data, we only need to adjust $T_{\text{scale}}$ during testing. 

With the trained neural network, we can derive the probabilistic mapping $\pi(\bar{\kappa}|\kappa)$ for each true parameter $\kappa$ in a continuous parameter space $\mathcal{K}$, which satisfies the individual-level privacy constrain. Next, we introduce the nonlinear transformation that converts the generated pseodo parameters $\bar{\kappa})$ to pseudo states that will be transmitted to CAVs.  


\subsection{Nonlinear Transformation}
\label{subsec: Nonlinear Transformation}
The nonlinear transformation module is a synthetic data generator that leverages the pseudo parameters to generate pseudo states $\bar{x}$ such that the 
joint probabilistic distribution function (p.d.f.) of $\bar{x}$ is ``convincing''. 
Specifically, let $\kappa = \left[\omega_{1},\omega_{2},\cdots,\omega_{r}\right]$ denote the true parameters of car-following model, and let $\bar{\kappa}=\left[\bar{\omega}_{1},\bar{\omega}_{2},\cdots,\bar{\omega}_{r}\right]$ 
denote the pseudo parameters chosen by the randomizer. The p.d.f. of the true states is represented by $p_{\kappa}\left(x_{1: T}\right)$ parametrized by $\kappa$. What we mean by ``convincing'' is that after the nonlinear transformation, the p.d.f. of pseudo states is $p_{\bar{\kappa}}\left(\bar{x}_{1: T}\right)$ parametrized by $\bar{\kappa}$ . In other words, we want to joint distribution of the pseudo states to be consistent with the assumed pseudo parameter $\bar{\kappa}$. 



For an HDV $i\in \Omega_\mathcal{H}$ equipped with a privacy filter, in the $t_{\text{s}}$-th step of the simulation, using the true parameters $\kappa_{i}$ and pseudo parameters $\bar{\kappa}_{i}$, we generate pseudo states using a recursive scheme according to:
\begin{equation}
\Gamma_{i}\rightarrow\left\{
\begin{aligned}
&\bar{v}_{i}({t_{\text{s}}})=F_{\bar{v}_{i}}^{-1}\left(d_1({t_{\text{s}}}) \mid \bar{s}_{i}({t_{\text{s}}}-1),\bar{v}_{i}({t_{\text{s}}}-1), v_{i-1}({t_{\text{s}}}-1), \bar{\kappa}_{i}\right),\\
&d_1({t_{\text{s}}})=F_{v_{i}}\left(v_{i}({t_{\text{s}}}) \mid s_{i}({t_{\text{s}}}-1), v_{i}({t_{\text{s}}}-1), v_{i-1}({t_{\text{s}}}-1), \kappa_{i}\right),\\
&\bar{s}_{i}({t_{\text{s}}})=F_{\bar{s}_{i}}^{-1}\left(d_2({t_{\text{s}}}) \mid \bar{s}_{i}({t_{\text{s}}}-1), \bar{v}_{i}({t_{\text{s}}}-1),\bar{v}_{i}({t_{\text{s}}}), v_{i-1}({t_{\text{s}}}-1),\bar{\kappa}_{i}\right),\\
&d_2(t_{\text{s}})=F_{s_{i}}\left(s_{i}({t_{\text{s}}}) \mid s_{i}({t_{\text{s}}}-1), v_{i}({t_{\text{s}}}-1), v_{i}({t_{\text{s}}}),v_{i-1}({t_{\text{s}}}-1), \kappa_{i}\right),\\
\end{aligned}
\right.
\label{eq:nonlinear transformation}
\end{equation}
\noindent where $v_{i-1}$ is the velocity of the preceding vehicle, and $F$ denotes the conditional cumulative distribution function (c.d.f)., i.e., $F_v\left(z \mid v_n, x_n, \kappa\right)=\int_{-\infty}^z p_{\kappa, v}\left(z \mid v_n, x_n\right) d z$. The velocity of the preceding vehicle is recognized as an external input of the system. The structure of the nonlinear transformation utilizes a feedforward-feedback scheme. The feedforward part computes the conditional c.d.f. $d_1,d_2$, while the feedback part computes the pseudo states $\bar{s}_{i},\bar{v}_{i}$ using the output of the randomizer.

\section{Analysis of Privacy-Utility Trade-off}  
\label{sec: Simulation Results}

This section presents numerical analyses to demonstrate the privacy-utility trade-off of the proposed framework. Section~\ref{subsec:simu_setting} describes the simulation setting. 
Section~\ref{subsec:privacy_analysis} performs privacy analysis for HDVs in different mixed-autonomy platoon scenarios as shown in Fig. \ref{fig: privacy scenarios}, whereby adversaries aim to infer the sensitive parameters of HDVs from the reported state information. 
Section~\ref{subsec:control_performance} quantifies the impact of the privacy filter on the accuracy of the reported states and the CAV control performance in terms of fuel consumption and average absolute velocity error.

\subsection{Simulation Setting} \label{subsec:simu_setting}
In this paper, following the setting of LCC \citep{wang2021leading}, we use the Full Velocity Difference Model (FVD) \citep{jiang2001full} as an example of  HDVs' car-following models:
\begin{equation}
    \mathbf{F} (\cdot)=\alpha_i\left(V\left(s_i(t)\right)-v_i(t)\right)+\beta_i (v_{i-1}(t)-v_{i}(t)),
\end{equation}
where $\alpha_i$ and $\beta_i$ are positive constants for HDV $i\in\Omega_H$ representing sensitivity coefficients. As stated in Assumption~\ref{asm:para_dist}, these sensitive parameters are assumed to follow an independent and identical probability distribution, which is publicly known. Specifically, for simplicity, both $\alpha_i$ and $\beta_i$ are assumed in this paper to follow a uniform distribution over the range $\left[0,1\right]$. Our proposed method can be easily applied to handle parameters with other theoretical and practical distributions (e.g., drawn from realistic vehicle trajectory datasets such as NGSIM). 
$V(s)$ represents the spacing-dependent desired velocity of human drivers with a specific form indicated below: 
\begin{equation}
V(s)= 
\begin{cases}
0, & s \leq s_{\mathrm{st}} \\
\frac{v_{\max }}{2}\left(1-\cos \left(\pi \frac{s-s_{\mathrm{st}}}{s_{\mathrm{go}}-s_{\mathrm{st}}}\right)\right), & s_{\mathrm{st}}<s<s_{\mathrm{go}} \\ 
v_{\mathrm{max}}, & s \geq s_{\mathrm{go}}
\end{cases}
\label{eq: range policy}
\end{equation}
where $s_{\mathrm{st}}$ and $s_{\mathrm{go}}$ represent two spacing thresholds, and $v_{\mathrm{max}}$ represents the maximum velocity. For simplicity of comparison, we assume that each vehicle has the same parameter values in $s_{\mathrm{st}}$, $s_{\mathrm{go}}$, and $v_{\mathrm{max}}$. 
Details for this can be found in \citet{jin2016optimal}.

\vspace{0.3em}\noindent \emph{Attacker settings.} We consider an attacker interested in learning the sensitive parameter for HDV $i$, i.e., $\kappa_i = \left[\alpha_i,\beta_i\right]$. 
Note that the attacker is not interested in learning $s_{\text{st}}$ and $s_{\text{go}}$ because the driving behavior (e.g., aggressiveness) is more intuitively reflected by parameters $\alpha$ and $\beta$. Other parameters associated with HDVs and CAVs in the experiments are fixed, as shown in Table~\ref{tab: parameters setup}.

\vspace{0.3em}\noindent \emph{Driving scenarios.} We evaluate the privacy-utility trade-off in two driving scenarios characterized by the movement of the head vehicle. The formation of the platoon consists of $5$ HDVs with indices $\Omega_\mathcal{H} = \{1,3,4,5,6\}$, $1$ CAV with index $\Omega_\mathcal{C} = \{2\}$, and $1$ head vehicle with index $0$.  \vspace{-0.6em}
\begin{itemize}
    \item  \emph{Sine-like velocity disturbance}: The head vehicle (HDV $0$) aims to maintain a specific target velocity but has a sine-like velocity disturbance with an amplitude of $3$ m/s. This scenario represents a typical platooning situation where the velocity of the head vehicle oscillates around a fixed value.\vspace{-0.6em}

    \item\emph{Emergency braking}: The head vehicle performs an emergency braking with a constant deceleration rate of $-3.5 \text{m/s}^2$ for $1.75$s before recovering to its equilibrium velocity with an acceleration of $5\text{m/s}^2$ for another $1.75$s. This scenario evaluates the performance of the proposed parameter privacy filter in extreme conditions. \vspace{-0.6em}
\end{itemize}

\begin{table}[ht] 
\centering
\renewcommand\arraystretch{1.2}
\caption{Values of key parameters used in case studies.}
\begin{tabular}{|c|ccccccccc|} 
\hline
Parameters & $s_{st}$ & $s_{go}$  & $a_{\min}$  & $a_{\max}$ & $\mu_{2,0}$ & $\eta_{2,0}$ & $\mu_{2,1}$ & $\eta_{2,1}$ &$\mu_{2,2}$\\
\hline
Values & 5 & 35 & -5 & 5 & 0 & 0.05 & -0.05 & 0.3 & 0.2\\
\hline
Parameters & $\eta_{2,2}$ & $\mu_{2,3}$ & $\eta_{2,3}$ & $\mu_{2,4}$ & $\eta_{2,4}$ & $\mu_{2,5}$ & $\eta_{2,5}$ & $\mu_{2,6}$ & $\eta_{2,6}$ \\
\hline
Values & -0.6 & -0.1 & 0.15 & -0.05 & 0.1 & -0.01 & 0.08 & -0.01 & 0.05\\
\hline
Parameters & $v^\star$ & $s^\star$ & $T_{\text{step}}$ & $T_{\text{ini}}$ & $w_v$ & $w_s$& $w_u$ & $\lambda_g$ & $\lambda_y$\\
\hline
Values & 15 & 20 & 0.01 & 0.2 & 1 & 0.5 & 0.1 & 1 & 1e3\\
\hline
\end{tabular}
\label{tab: parameters setup}
\vspace{-0.1cm}
\end{table}

\subsection{Performance of Privacy Preservation} \label{subsec:privacy_analysis}

In this subsection, we evaluate the privacy protection performance of the parameter privacy filter under the aforementioned two driving scenarios. Specifically, we assume that an adversary with access to the shared error states attempts to infer the sensitive car-following parameters of HDV $4$ (hereafter also referred to as the ego vehicle). 
The true car-following parameters of HDV $4$ is $\kappa = \left[0.6,0.9\right]$. 

\begin{figure}[ht]
    \centering
    \subcaptionbox{}{\includegraphics[width=6cm]{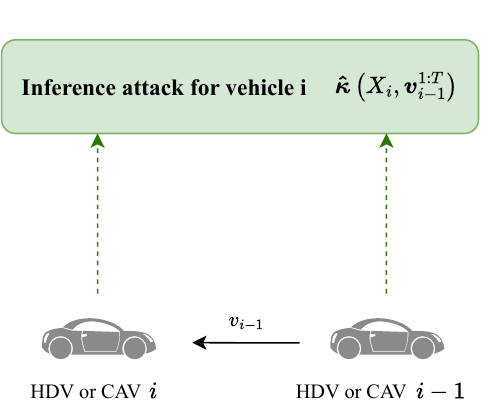}}
    \subcaptionbox{}{\includegraphics[width=6cm]{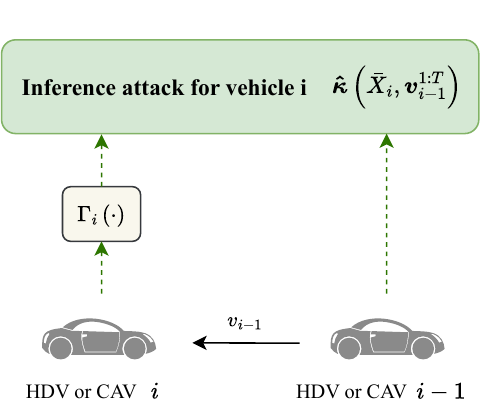}}
    \subcaptionbox{}{\includegraphics[width=6cm]{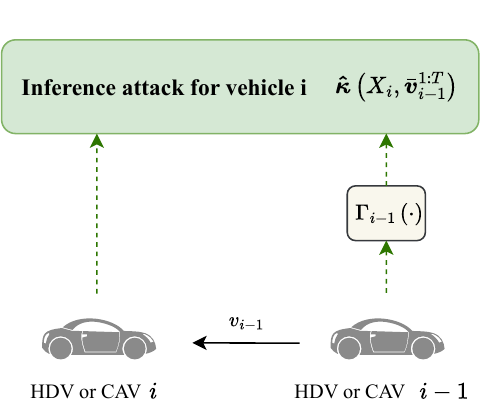}}
    \subcaptionbox{}{\includegraphics[width=6cm]{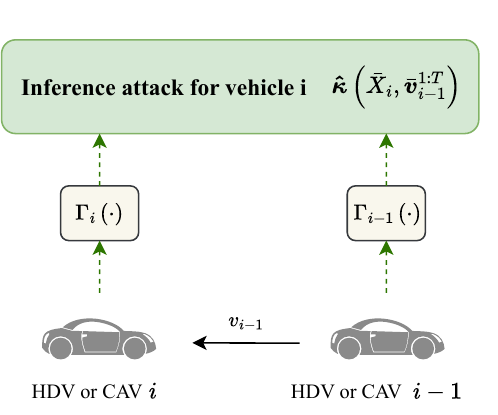}}
    
    \caption{Implementation scenarios of the parameter privacy filter in platoon control: (a) neither vehicle $i$ nor its preceding vehicle $i-1$ has implemented the privacy filter, (b) only vehicle $i$ has implemented the privacy filter, (c) only vehicle $i-1$ has implemented the privacy filter, and (d) both vehicles have implemented the privacy filter.  }
    \label{fig: privacy scenarios}
\end{figure}

\emph{Implementation Scenarios}. We investigate different implementation scenarios of the parameter privacy filter, as shown in Fig. \ref{fig: privacy scenarios}. The scenario in Fig.~\ref{fig: privacy scenarios} (a) represents a baseline where neither the preceding (i.e., HDV $3$) nor the ego vehicle (i.e., HDV $4$) employs a privacy filter, leaving the HDV parameters exposed and unshielded. In contrast, the scenario shown in Fig. \ref{fig: privacy scenarios} (b) illustrates the case where the ego vehicle utilizes a privacy filter, actively safeguarding its own parameter privacy. The scenario shown in Fig.\ref{fig: privacy scenarios} (c) evaluates the effect of equipping the preceding vehicle with a privacy filter, which, due to the system's interconnectivity, inadvertently enhances the parameter privacy of the ego vehicle as well. The scenario shown in Fig.\ref{fig: privacy scenarios}  (d) presents a scenario where both the preceding and the ego vehicles are equipped with privacy filters.

\emph{Attacker Model}. We employ an attacker model to quantify the level of privacy protection in these scenarios. The attackers in the simulation (i.e., any adversary who can access the transmitted data of an HDV to the central unit) are assumed to use the nonlinear least squares method~\citep{teunissen1990nonlinear} to estimate the sensitive parameters of HDVs from the reported error states. We use this method as the attacker model because it is a classic and effective approach for nonlinear system identification. Note that since HDVs report a sequence of distorted error states instead of distorted raw states, the attacker needs to simultaneously estimate the sensitive parameter $\kappa_i$ and the equilibrium state $s_i^\star$. To this end, the attacker can add a constraint to the attacker model, i.e., $\mathbf{F} _{\mathbb{\kappa}_i}\left(s_i^\star, v^\star, v^\star\right)=0$, whereby the equilibrium velocity $v^\star$ can be easily obtained from the head vehicle. 
Specifically, for FVM, this constraint can be represented by $s_i^\star = \frac{s_{\mathrm{go}}-s_{\mathrm{st}}}{\pi}\arccos\left(1-\frac{2v^{\star}}{v_{\max}}\right)+s_{\mathrm{st}}~\forall i \in \Omega_{\mathcal{H}}$. 

\emph{Benchmarks}. With the attacker model, we characterize the privacy protection performance by the attackers' estimation accuracy of the sensitive parameters, i.e., normalized Root Mean Squared Error (RMSE). We evaluate such a performance index in the aforementioned privacy scenarios by comparing the following three benchmark methods with various levels of privacy considerations:\vspace{-0.3em} 
\begin{itemize}
    \item The \emph{proposed parameter privacy filter} with various levels of $I_0$;  \vspace{-0.3em}
    \item A \emph{noise-adding mechanism} with additive Gaussian random noise, which enhances privacy by introducing stochasticity to the original state sequence (as has been done in existing literature); \vspace{-0.3em}
    \item \emph{No privacy considerations}. \vspace{-0.3em}
\end{itemize}
The comparison between these three benchmark methods demonstrates the value of the parameter privacy filter in protecting sensitive parameters in car-following models. 

Our evaluation further encompasses the parameter privacy performance of the two approaches for solving the optimization problem associated with the randomizer: (i) the discretization approach with the aggregated-level privacy preservation constraint in Section~\ref{subsubsec: Discrete Parameter Privacy Filter for HDVs}, and (ii) the learning-based randomizer approach with the individual-level privacy preservation constraint in Section~\ref{subsubsec: Continuous Parameter Privacy Filter for HDVs}. 

\subsubsection{Parameter Privacy Filter with Discretization-Based Randomizer}
\label{subsubsec: Discrete Parameter Privacy Filter for HDVs}
We first present the results of the discretization-based approach for solving the optimization problem associated with the randomizer. Specifically, we divide the parameter space $[0,1]\times[0,1]$ into 5 regions and assume that the true parameters for the $5$ HDVs take values in the following discrete set with equal probability:
\begin{equation*}
\mathcal{D}=\left\{\left[\begin{array}{c}
0.6 \\
0.9
\end{array}\right],\left[\begin{array}{l}
0.4 \\
0.7
\end{array}\right],\left[\begin{array}{l}
0.8 \\
0.5
\end{array}\right],\left[\begin{array}{l}
1.0 \\
0.8
\end{array}\right],\left[\begin{array}{l}
0.2 \\
0.2
\end{array}\right]\right\}.
\end{equation*}

The pseudo parameters are selected from the following set by solving the randomizer Eq.~\eqref{eq:opt-randomizer-discrete}:
\begin{equation*}
\bar{\mathcal{K}}=\left\{\left[\begin{array}{c}
0.5 \\
0.8
\end{array}\right],\left[\begin{array}{l}
0.2 \\
0.4
\end{array}\right]\right\}.
\end{equation*}

\begin{figure}[ht]
         \centering
                 \subcaptionbox{}{
            \includegraphics[width=5.1cm]{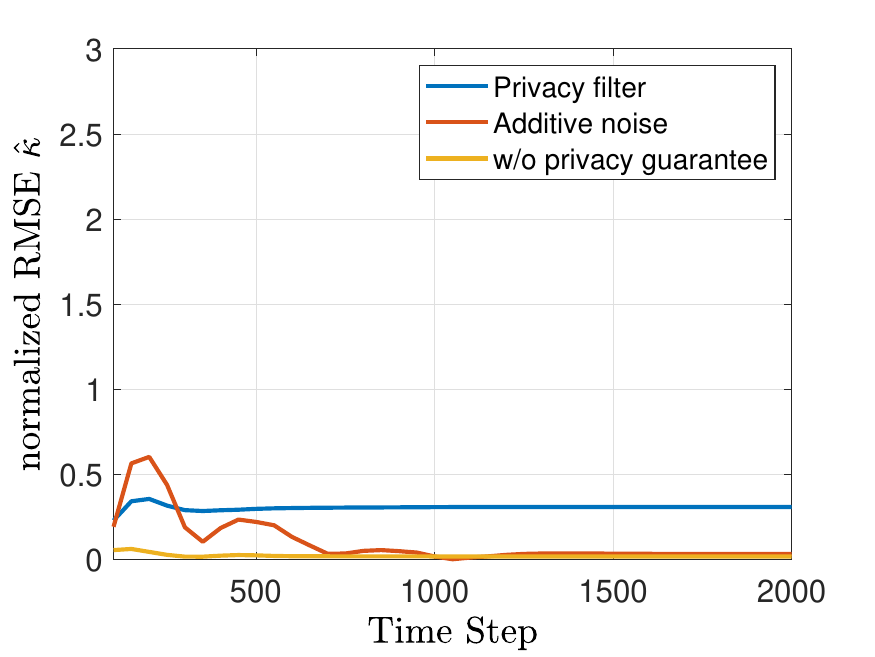}
        }
         \subcaptionbox{}{
            \includegraphics[width=5.1cm]{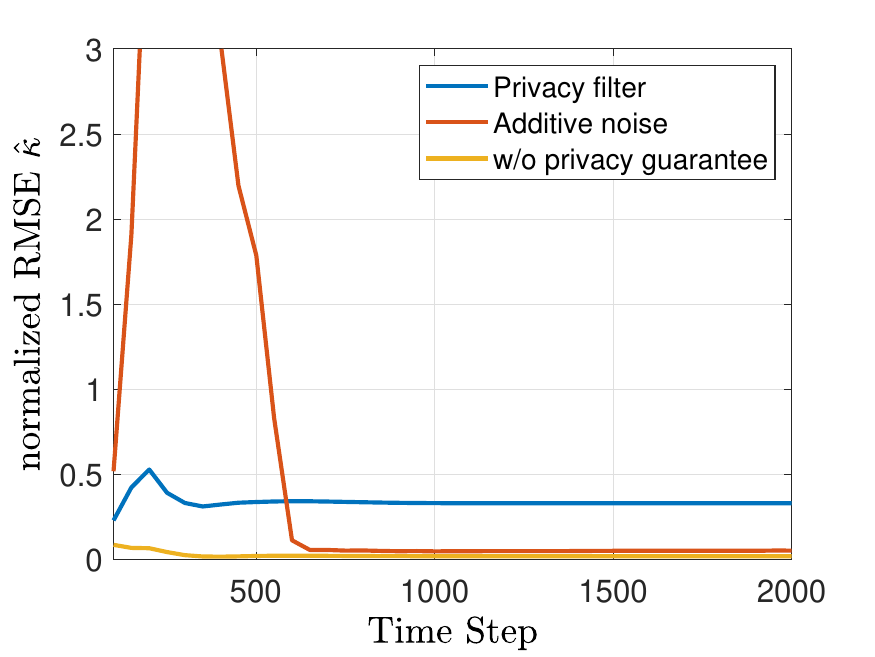}
        }\\
        \subcaptionbox{}{
            \includegraphics[width=5.1cm]{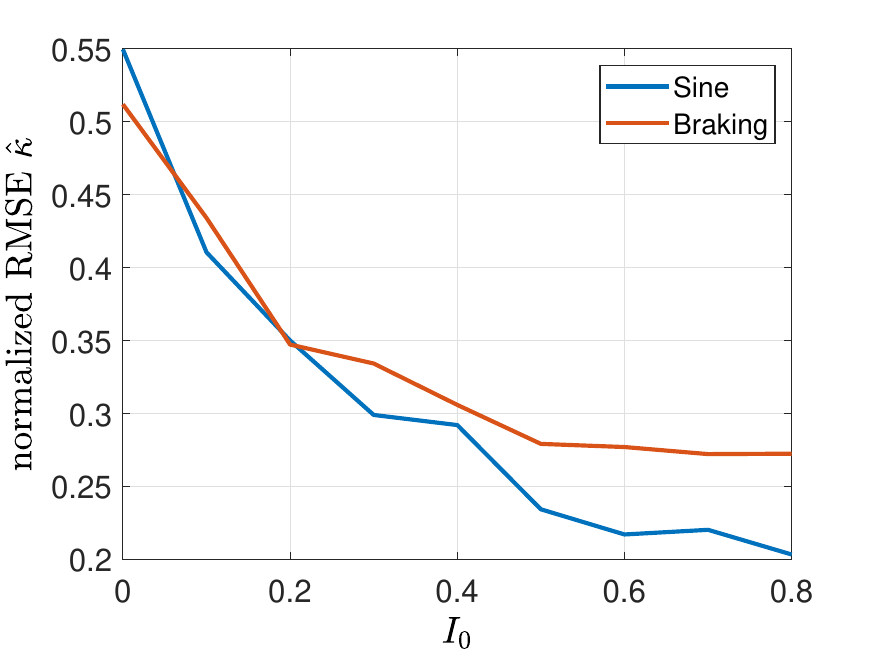}
        }
        \subcaptionbox{}{
            \includegraphics[width=5.1cm]{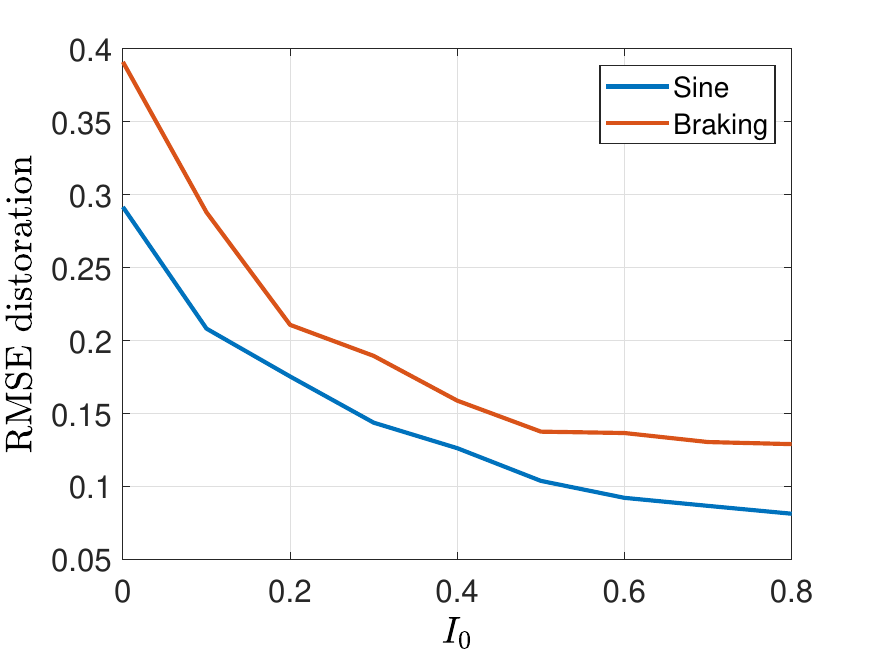}
        }
        \caption{Simulation results for the scenario in Fig.~\ref{fig: privacy scenarios} (b) using parameter privacy filter with a discretization-based randomizer. (a) and (b) show the normalized RMSE of the parameter estimator over time for the driving scenarios of sine-like velocity disturbances and emergency braking, respectively, with $I_0=0.2$ and a Gaussian additive noise intensity of $0.2$. Note that the higher the RMSE, the better the privacy protection. 
        (c) illustrates the normalized RMSE error of the parameter estimator in scenarios with various levels of information budget $I_0$ for two driving scenarios. 
        (d) represents average total distortion in scenarios with various levels of information budget $I_0$ for two driving scenarios. }
        \label{fig: performance of privacy preservation HDVs scenario 1}
\end{figure}

\begin{figure}[ht]
         \centering         
        \subcaptionbox{}{
            \includegraphics[width=5.1cm]{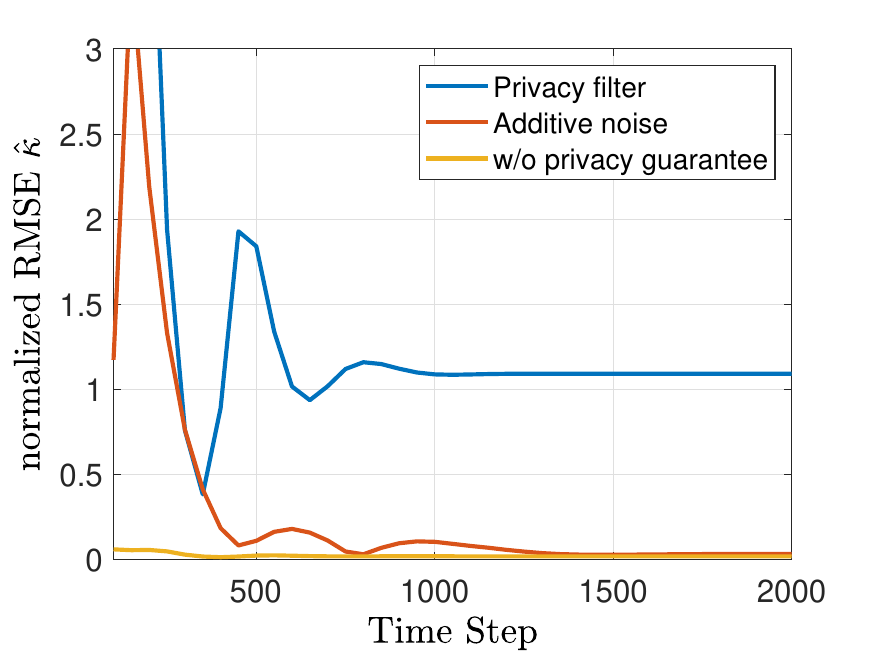}
        }
        \subcaptionbox{}{
            \includegraphics[width=5.1cm]{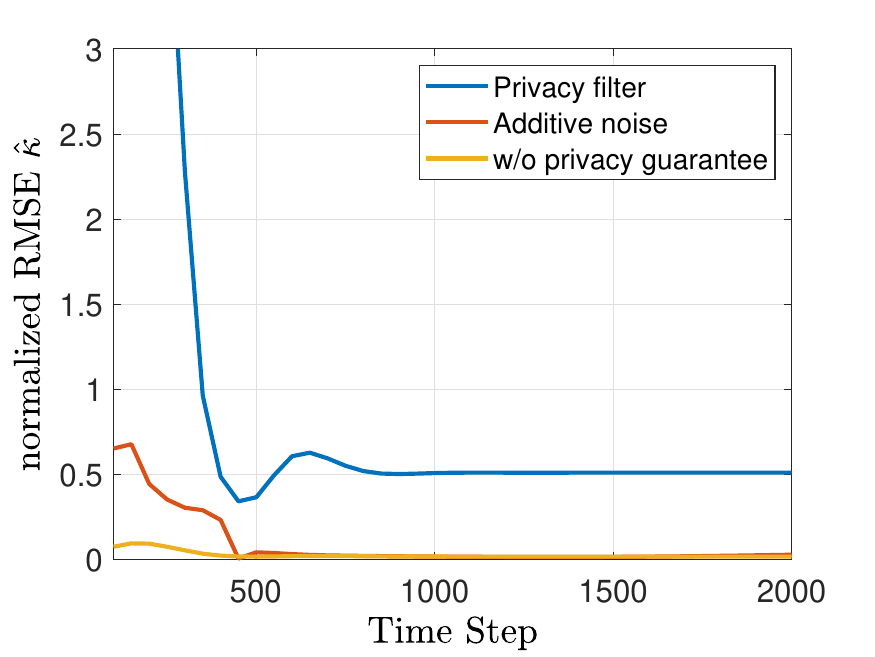}
        }
        \subcaptionbox{}{
            \includegraphics[width=5.1cm]{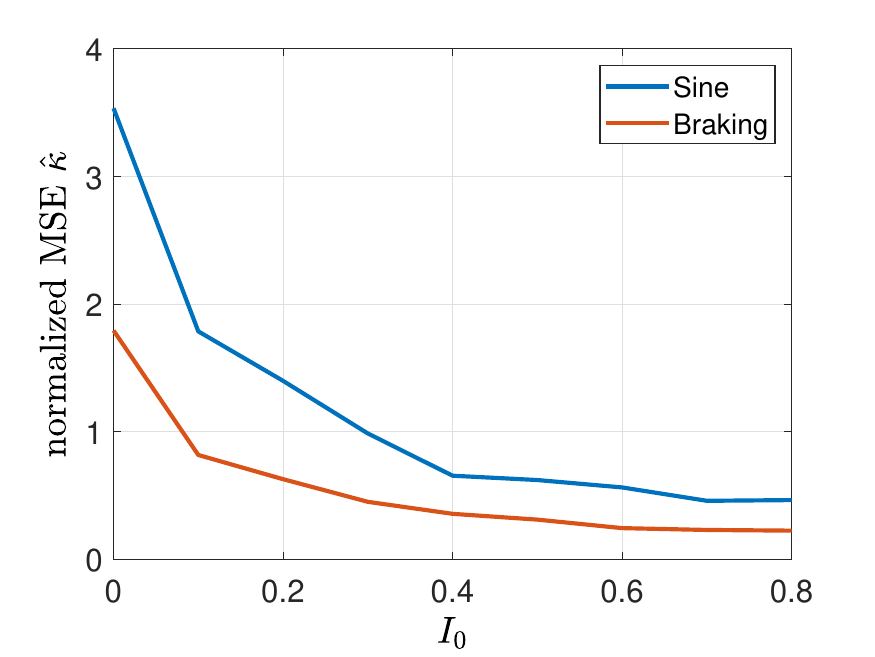}
        }
        \caption{Simulation results for the scenario in Fig.~\ref{fig: privacy scenarios} (c) using the parameter privacy filter with a discretization-based randomizer. (a) and (b) indicate the normalized RMSE of the parameter estimator as a function of time for the driving scenarios of sine-like velocity disturbances and emergency braking, respectively, with $I_0=0.2$ and a Gaussian additive noise intensity of $0.2$. (c) shows the normalized RMSE error of the parameter estimator versus the level of information leakage $I_0$ for two driving scenarios. }
        \label{fig: performance of privacy preservation HDVs scenario 2}
\end{figure}

\begin{figure}[ht]
         \centering
         \subcaptionbox{}{
            \includegraphics[width=5.1cm]{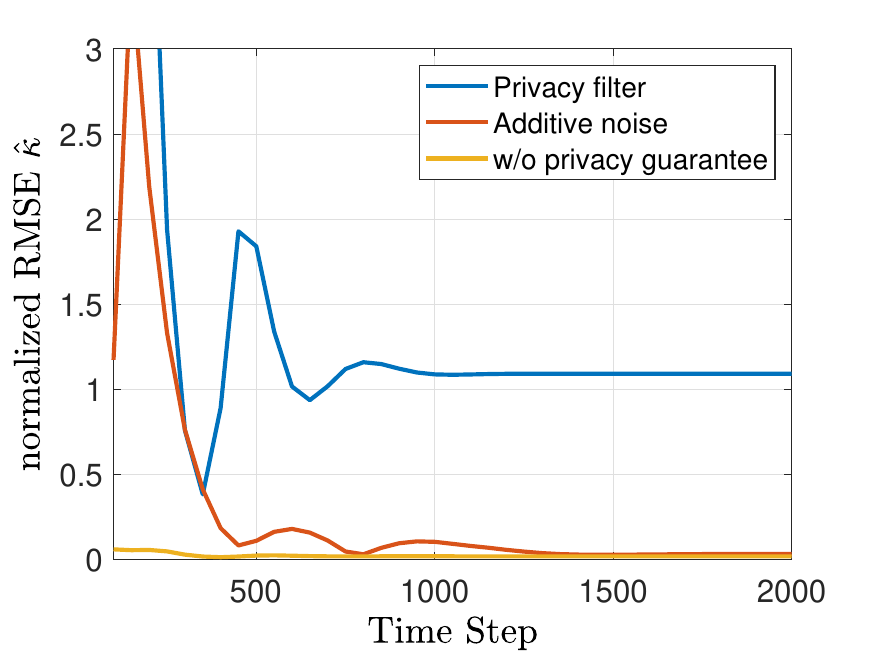}
        }        
         \subcaptionbox{}{
            \includegraphics[width=5.1cm]{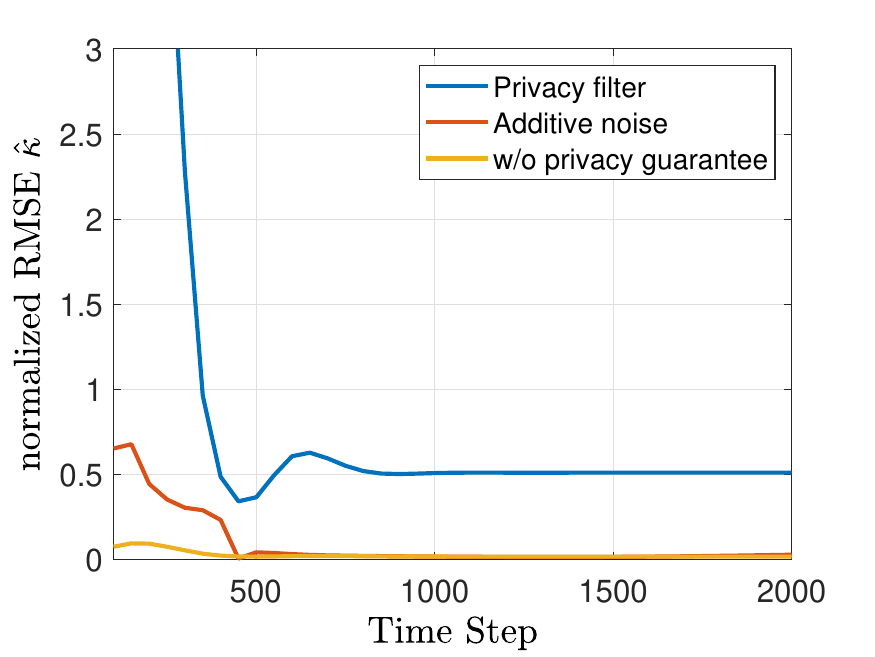}
        }\\    
        \subcaptionbox{}{
            \includegraphics[width=5.1cm]{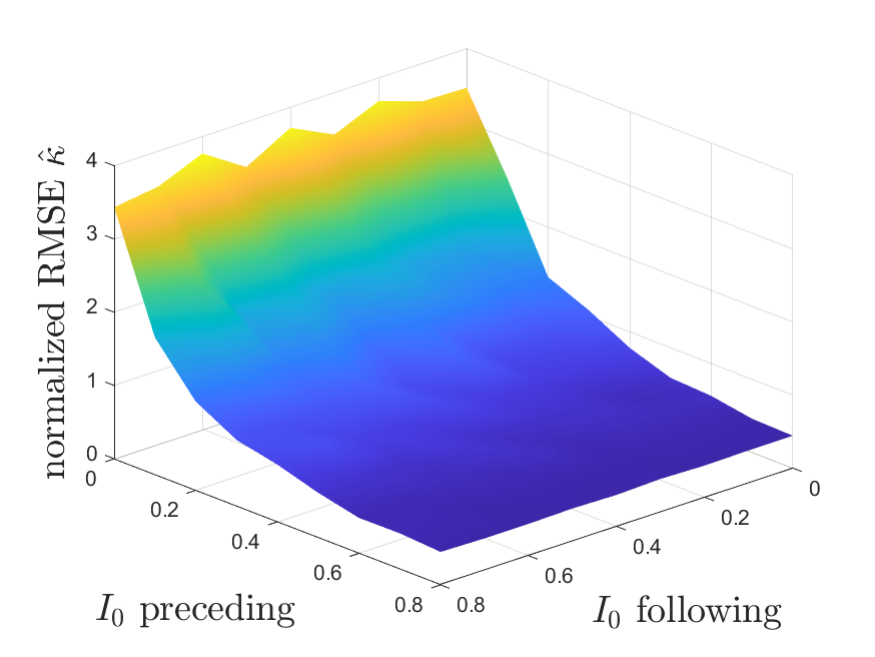}
        }        
         \subcaptionbox{}{
            \includegraphics[width=5.1cm]{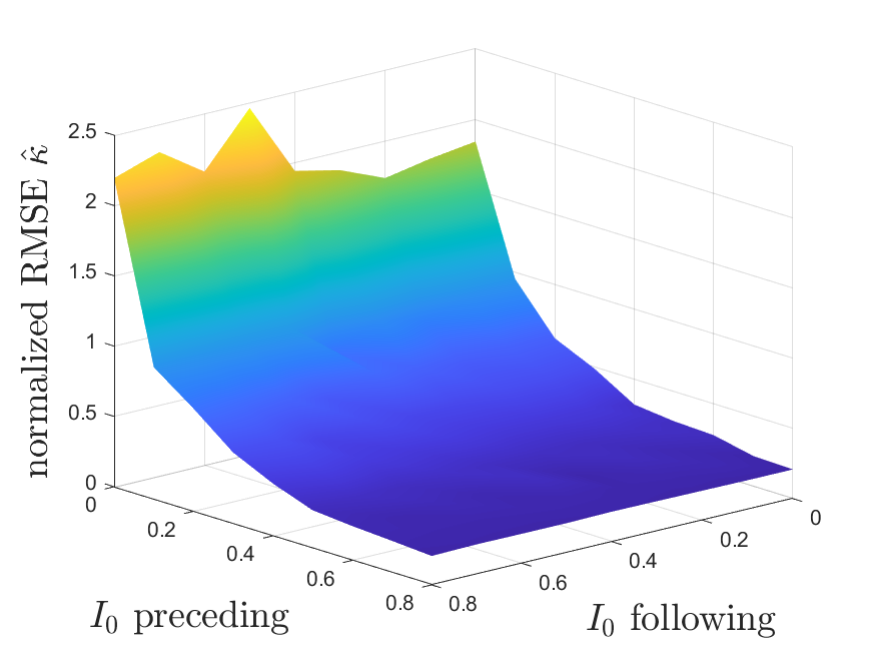}
        }
        \caption{Simulation results for the scenario in Fig.~\ref{fig: privacy scenarios} (d) using the parameter privacy filter with a discretization-based randomizer. (a) and (b) indicate normalized RMSE of the parameter estimator as a function of time for the two driving scenarios of sine-like velocity disturbances and emergency braking, respectively, where $I_0$ is set to be $0.2$ for both the preceding and following vehicles, and the intensity of the Gaussian additive noise is $0.2$. (c) and (d) indicate the relationship between the normalized RMSE error of the parameter estimator (z-axis) and the privacy levels for the preceding and following vehicles (x-axis and y-axis). }
        \label{fig: performance of privacy preservation HDVs scenario 3}
\end{figure}

Fig.~\ref{fig: performance of privacy preservation HDVs scenario 1}, Fig. \ref{fig: performance of privacy preservation HDVs scenario 2} and Fig. \ref{fig: performance of privacy preservation HDVs scenario 3} show the simulation results to evaluate the performance of privacy protection of the three implementation scenarios as illustrated in Fig.~\ref{fig: privacy scenarios}\,(b)-Fig.~\ref{fig: privacy scenarios}\,(d), respectively. Overall, the results demonstrate that the parameter privacy filter consistently performs effectively in all three scenarios, with the trade-off between privacy and utility being influenced by the selected privacy level $I_0$.\vspace{0.5em}

\noindent \emph{Value of the parameter privacy filter}. As illustrated in Fig.~\ref{fig: performance of privacy preservation HDVs scenario 1}\,(a) and Fig.~\ref{fig: performance of privacy preservation HDVs scenario 1}\,(b), we compare the parameter estimation error of the three methods mentioned above: the parameter privacy filter, noise-adding mechanism with additive Gaussian noise, and no privacy protection under two driving scenarios. The privacy level of the privacy filter, $I_0$, is set to $0.2$, and the mean of the Gaussian additive noise is set to $0.2$ to ensure similar distortion levels of the system states for these two methods. The results indicate that the RMSE of $\hat{\kappa}$ converges after approximately $700$ steps for all three methods. Furthermore, the parameter privacy filter can prevent adversaries from accurately reconstructing the sensitive parameters from the shared data, as the normalized RMSE for this method converges to a relatively high value of $0.4$. On the other hand, both of the other two methods fail to protect these sensitive parameters, as the adversaries can eventually accurately infer these parameters. This is expected as the noise added by the noise-adding mechanism can be easily smoothed by any filtering algorithm. This shows that the parameter privacy filter can effectively protect the sensitive parameters of HDVs' car-following models. \vspace{0.5em}

\noindent \emph{Impact of the privacy level $I_0$}. As illustrated in Fig.~\ref{fig: performance of privacy preservation HDVs scenario 1} (c) and (d), the normalized RMSE of $\hat{\kappa}$ and the state distortion decreases as $I_0$ increases for both driving scenarios. This is because the privacy bound of the randomizer will be relaxed as $I_0$ increases, leading to lower distortions. Note that the relation between these two values and $I_0$ may not appear to be strictly decreasing because the choice of $\bar{\kappa}$ is random. \vspace{0.5em}

\noindent \emph{Impact of implementing the parameter privacy filter on the preceding vehicle}. We demonstrate that vehicle privacy can be preserved by applying the privacy filter for the preceding vehicle due to the interconnected structure of the platoon system. For example, HDV $3$ is equipped with a privacy filter, while there is no privacy preservation for HDV $4$. Then, the adversary receives the pseudo velocity $\bar{v}^{\text{err}}_{3}$ rather than the true velocity $v^{\text{err}}_{3}$ of HDV 3. This can substantially impact any inference attacks for the HDV $4$ even though it is not equipped with a privacy filter because the preceding vehicle's velocity is crucial for estimating the control gain or car-following parameter of the following vehicle. This conclusion can be validated by the results shown in Fig. \ref{fig: performance of privacy preservation HDVs scenario 2}, where a predetermined privacy level $I_0=0.2$ is used. In fact, the level of protection is even better if the parameter filter is implemented in the preceding vehicle because the external input (i.e., $v_{i-1}$) significantly influences the estimation accuracy of the attacker. 
Moreover, in Fig. \ref{fig: performance of privacy preservation HDVs scenario 3} (a) and (b), it is evident that if 
both the preceding and the ego vehicles implement the parameter privacy filter, the sensitive parameters of the ego vehicle can be effectively protected. In particular, it is shown in Fig. \ref{fig: performance of privacy preservation HDVs scenario 3}\,(c) and (d) that the effectiveness of privacy preservation is influenced by both the privacy level of the privacy filter installed in the preceding vehicle and the privacy level of the following vehicle itself.

\subsubsection{Parameter Privacy Filter with a Learning-Based Randomizer}
\label{subsubsec: Continuous Parameter Privacy Filter for HDVs}
As $\alpha$ and $\beta$ are assumed to be uniformly distributed over the range $\left[0,1\right]$, we generate training dataset based on a true parameter set $\mathcal{D} = \left\{\left[\alpha,\beta\right]| \alpha \in \mathcal{M},\beta \in \mathcal{M}\right\}$ with $\mathcal{M}=\left\{0.1,0.325, 0.55, 0.775, 1\right\}$. 
We select the pseudo parameter set $\bar{\mathcal{K}}$ to be the same as the discretization-based approach. For the neural network-based randomizer, we utilized a fully connected neural network with an input size of 2 (dimension of $\kappa$), two hidden layers comprising 200 and 300 neurons, respectively, and an output layer of size 2 (cardinality of $\bar{\mathcal{K}}$).

With the integration of the individual-level privacy preservation constraint and the neural network-based randomizer, we have the following results for the three implementation scenarios of the privacy filter illustrated in Figs.~\ref{fig: privacy scenarios} (b)-(d). \vspace{0.5em}

\begin{figure}[ht]
         \centering
         \subcaptionbox{}{
            \includegraphics[width=5.1cm]{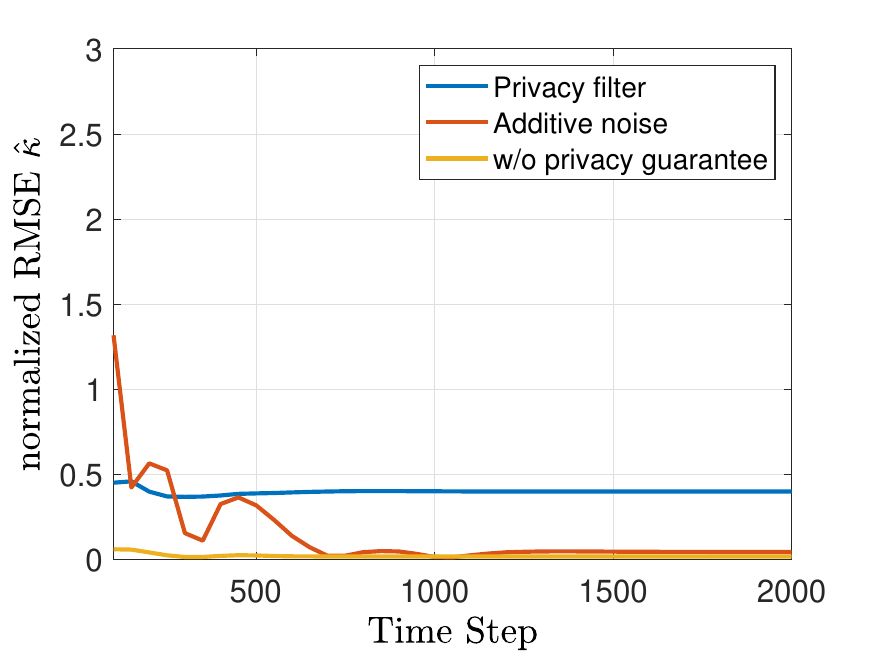}
        }
        \subcaptionbox{}{
            \includegraphics[width=5.1cm]{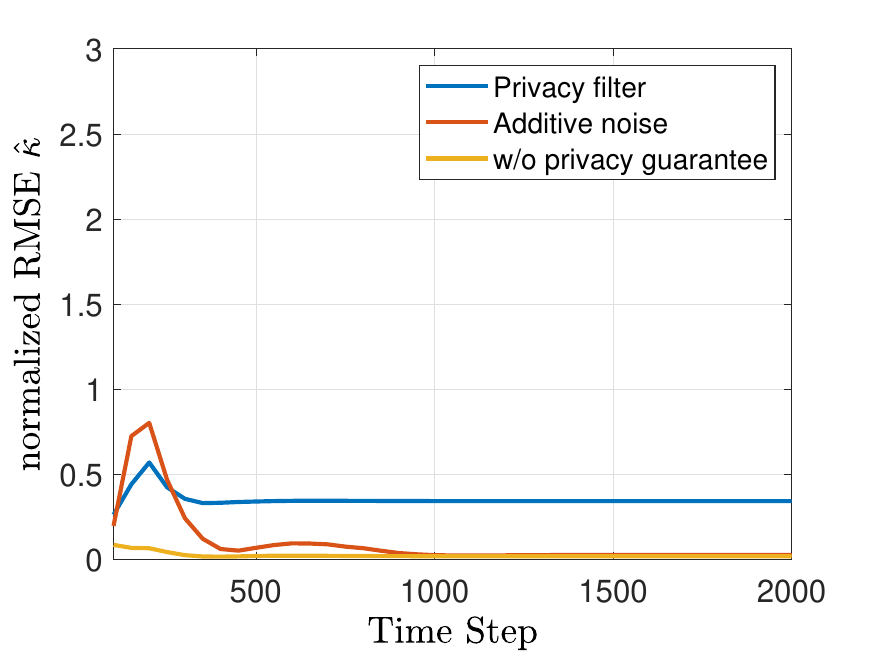}
        }\\
        \subcaptionbox{}{
            \includegraphics[width=5.1cm]{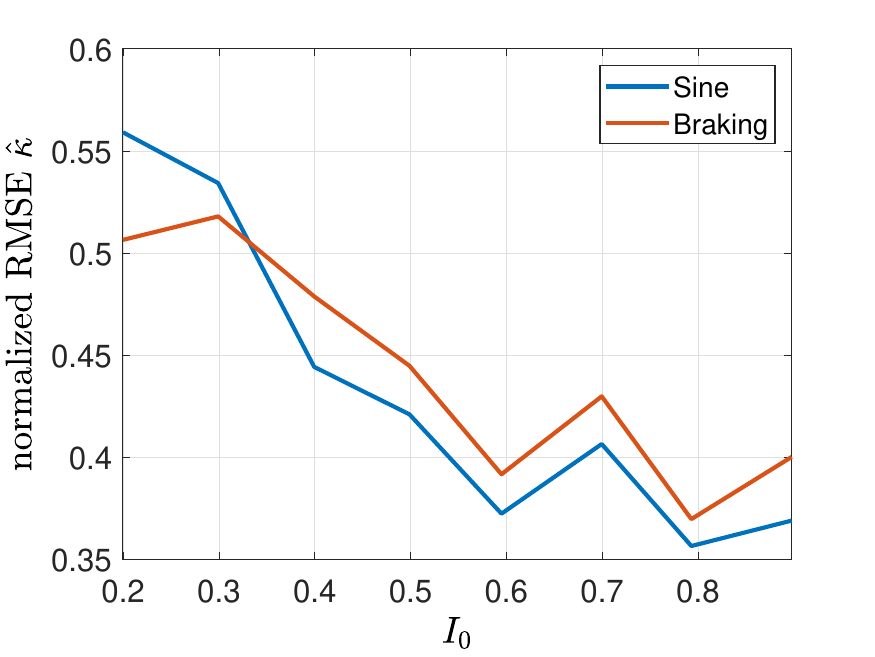}
        }
        \subcaptionbox{}{
            \includegraphics[width=5.1cm]{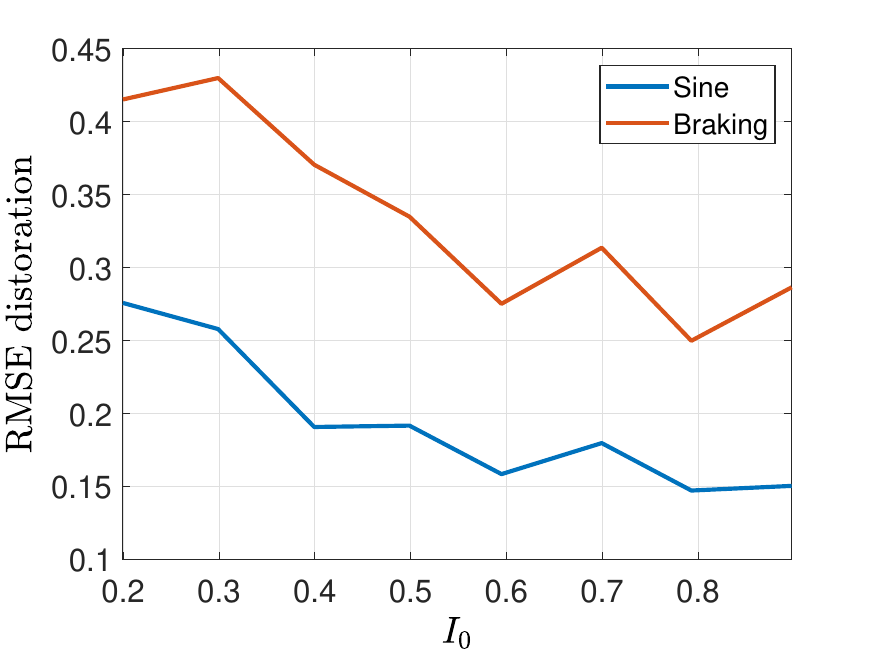}
        }
        \caption{Simulation results for the scenario in Fig.~\ref{fig: privacy scenarios} (b) using continuous parameter privacy filter. (a) and (b) show the normalized RMSE of the parameter estimator over time for driving scenarios involving sine-like velocity disturbances and emergency braking, respectively, with $I_0=0.2$ and a Gaussian additive noise intensity of $0.2$.
        (c) illustrates the normalized RMSE error of the parameter estimator versus the level of information leakage $I_0$ for two driving scenarios. 
        (d) represents average total distortion versus the level of information leakage $I_0$ for two driving scenarios.}
        \label{fig: continuous privacy filter performance for HDVs scenario 1}
\end{figure}

\begin{figure}[ht]
         \centering
        \subcaptionbox{}{
            \includegraphics[width=5.1cm]{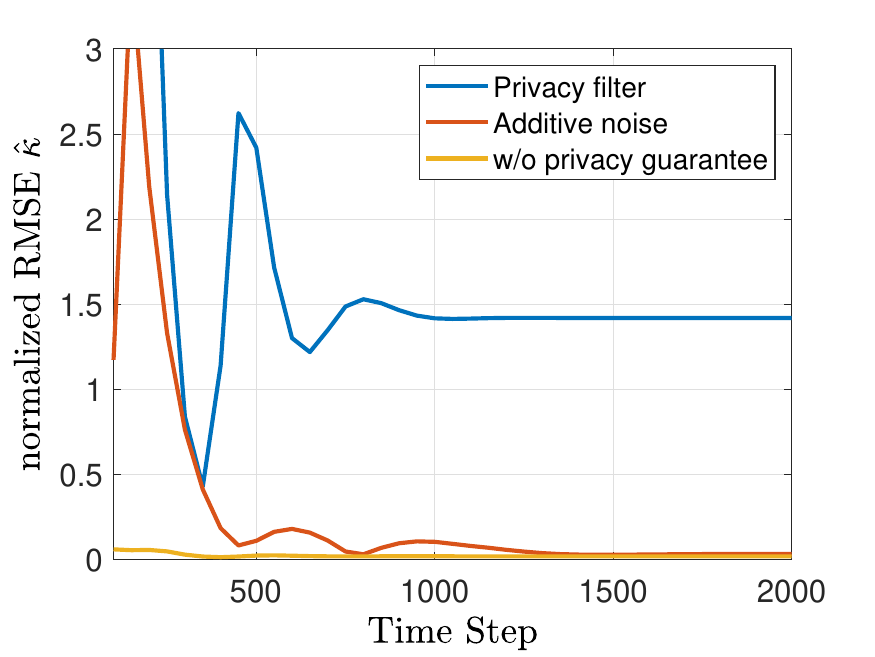}
        }         
         \subcaptionbox{}{
            \includegraphics[width=5.1cm]{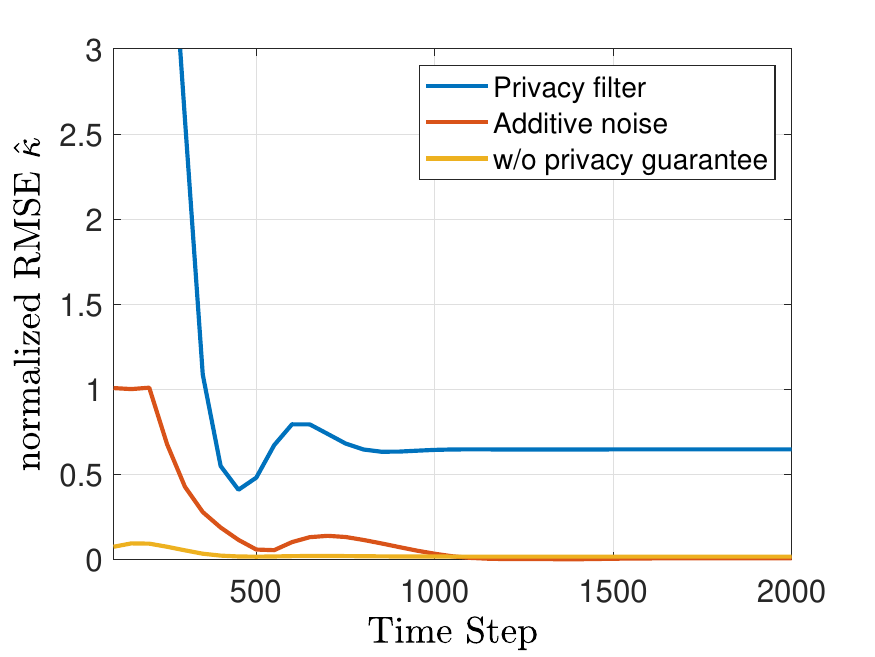}
        }    
        \subcaptionbox{}{
            \includegraphics[width=5.1cm]{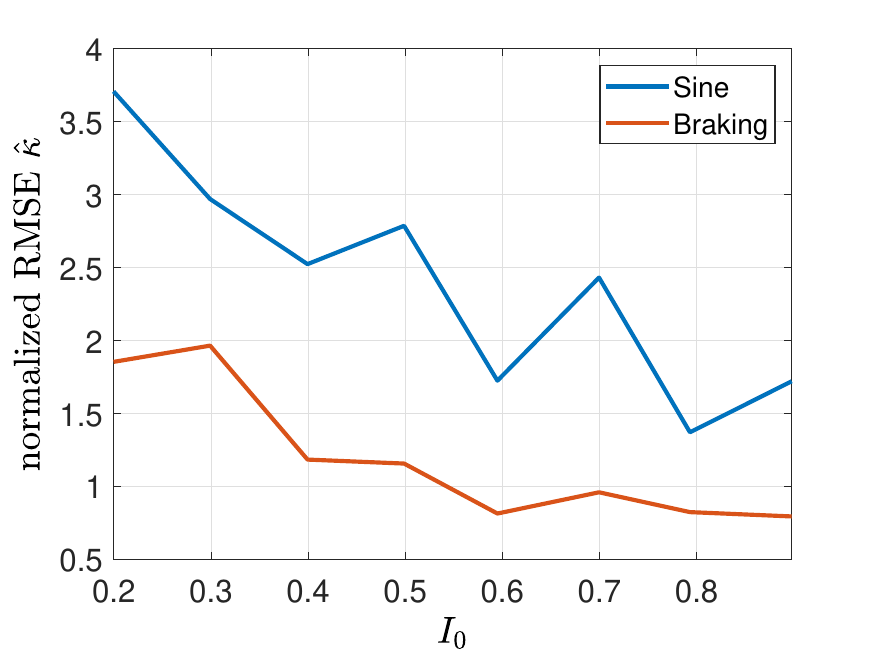}
        }
        \caption{Simulation results for the scenario in Fig.~\ref{fig: privacy scenarios} (c) using continuous parameter privacy filter. (a) and (b) indicate the normalized RMSE of the parameter estimator as a function of time for the two driving scenarios when $I_0=0.2$, the intensity of the Gaussian additive noise is $0.2$. (c) shows the normalized RMSE error of the parameter estimator versus the level of information leakage $I_0$ for two driving scenarios. }
        \label{fig: continuous privacy filter performance for HDVs scenario 2}
\end{figure}

\begin{figure}[ht]
         \centering
         \subcaptionbox{}{
            \includegraphics[width=5.1cm]{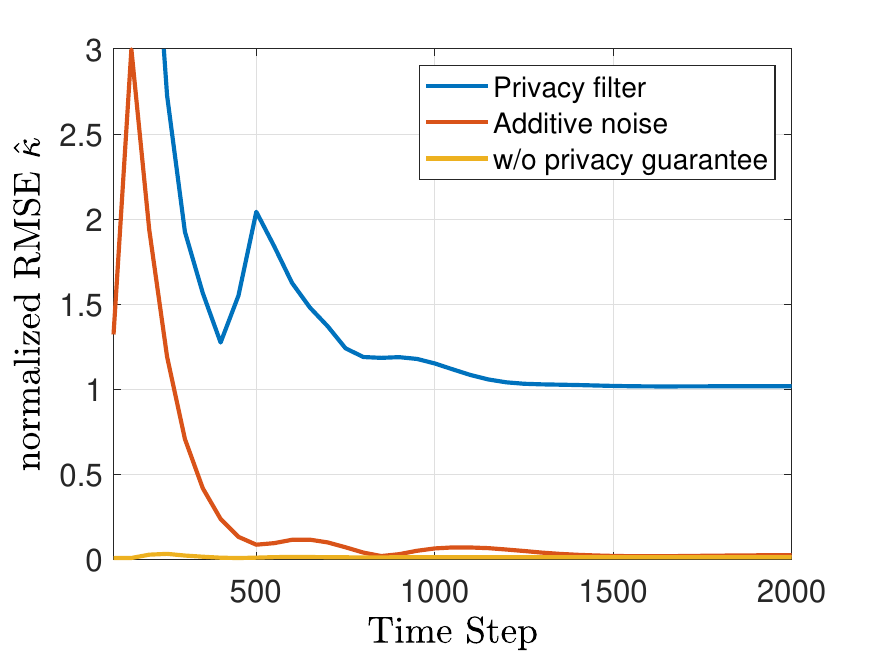}
        }
        \subcaptionbox{}{
            \includegraphics[width=5.1cm]{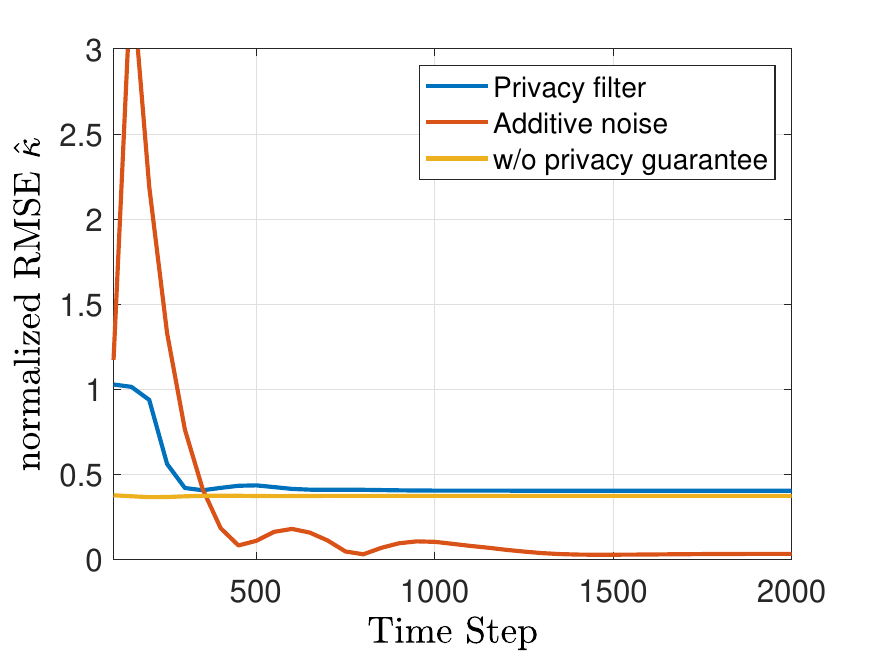}
        }\\
        \subcaptionbox{}{
            \includegraphics[width=5.1cm]{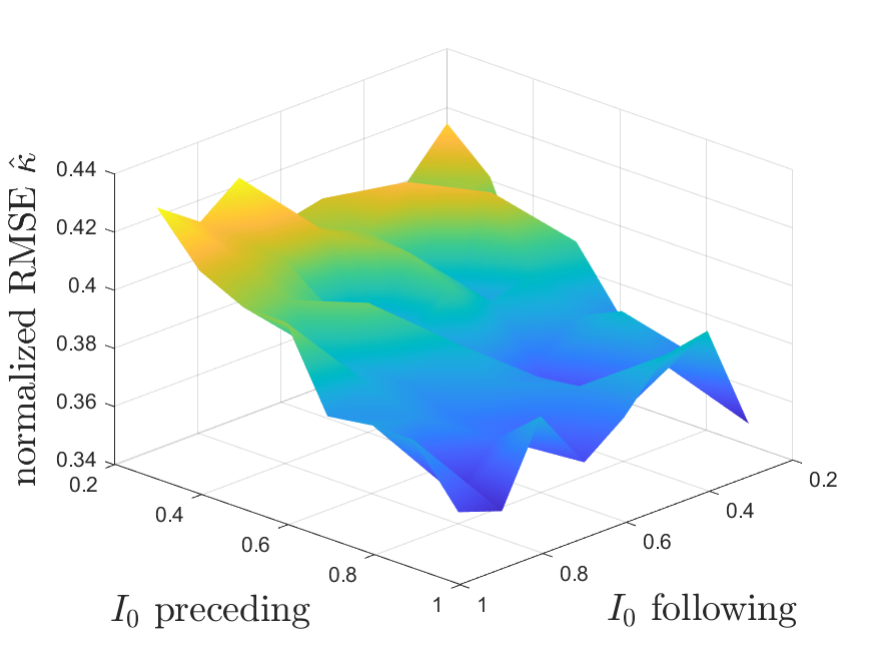}
        }
        \subcaptionbox{}{
            \includegraphics[width=5.1cm]{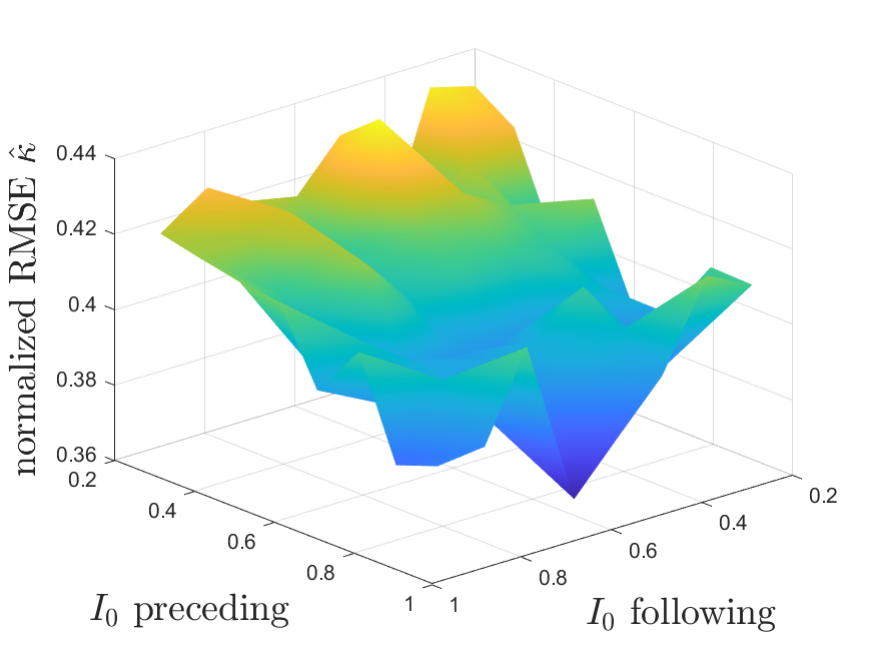}
        }
        \caption{Simulation results for the scenario in Fig.~\ref{fig: privacy scenarios} (d) using continuous parameter privacy filter. (a) and (b) indicate normalized RMSE of the parameter estimator as a function of time for the two driving scenarios when $I_0\ \text{preceding}=0.2$ and $I_0\ \text{following}=0.2$, the intensity of the Gaussian additive noise is $0.2$. (c) and (d) indicate the relationship between the normalized RMSE error of the parameter estimator (z-axis) and the privacy levels for the preceding and following vehicles (x-axis and y-axis). }
        \label{fig: continuous privacy filter performance for HDVs scenario 3}
\end{figure}

\noindent \emph{Value of the parameter privacy filter}. The results of the three implementation scenarios with different levels of privacy considerations are illustrated in Fig. \ref{fig: continuous privacy filter performance for HDVs scenario 1}, Fig. \ref{fig: continuous privacy filter performance for HDVs scenario 2}, and Fig. \ref{fig: continuous privacy filter performance for HDVs scenario 3}, which show that the privacy-preserving performance of the privacy filter aligns with that of the discretization approach. These results effectively demonstrate the effectiveness of the proposed neural network-based randomizer and the individual-level privacy preservation constraints. 

\begin{figure}[ht]
         \centering
         \subcaptionbox{$I_0=0.4$}{
            \includegraphics[width=5.1cm]{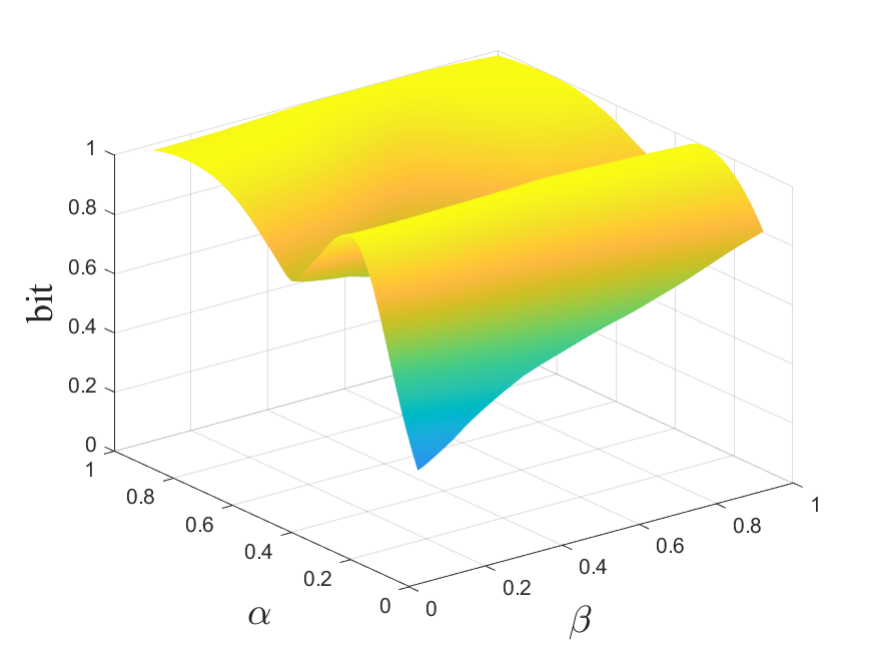}
        }
        \subcaptionbox{$I_0=0.6$}{
            \includegraphics[width=5.1cm]{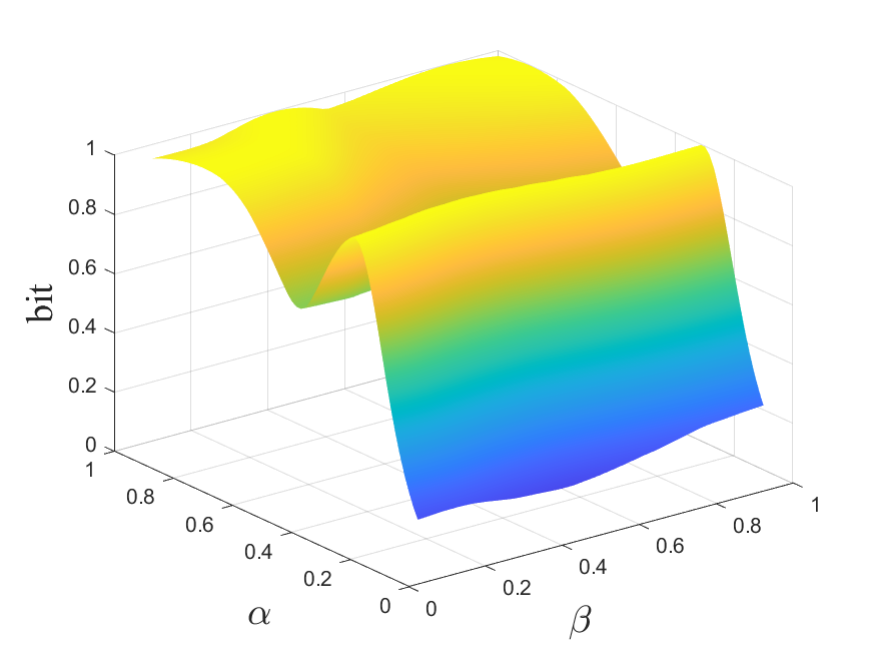}
        }
        \subcaptionbox{$I_0=0.8$}{
            \includegraphics[width=5.1cm]{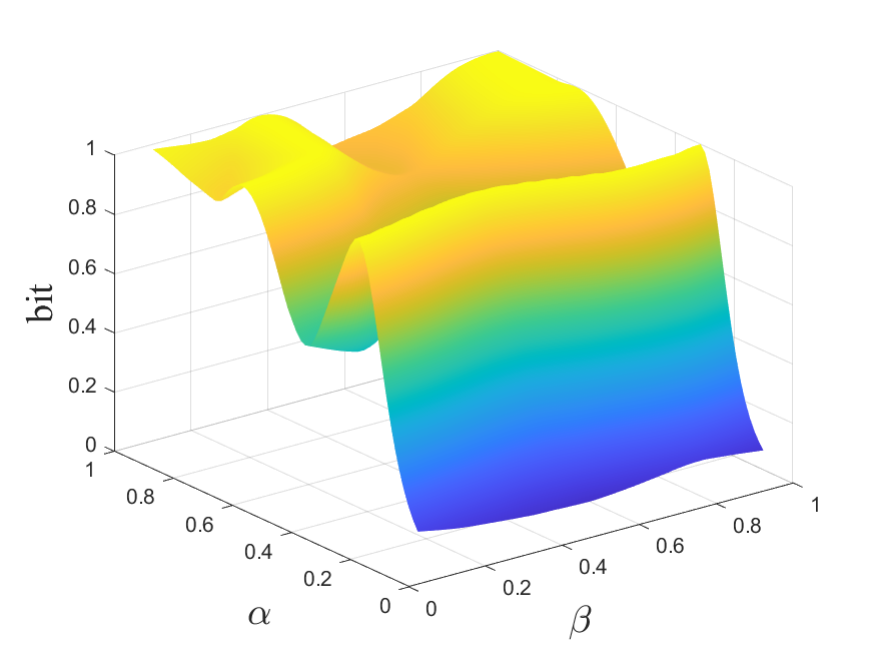}
        }
         \subcaptionbox{$I_0=0.4$}{
            \includegraphics[width=5.1cm]{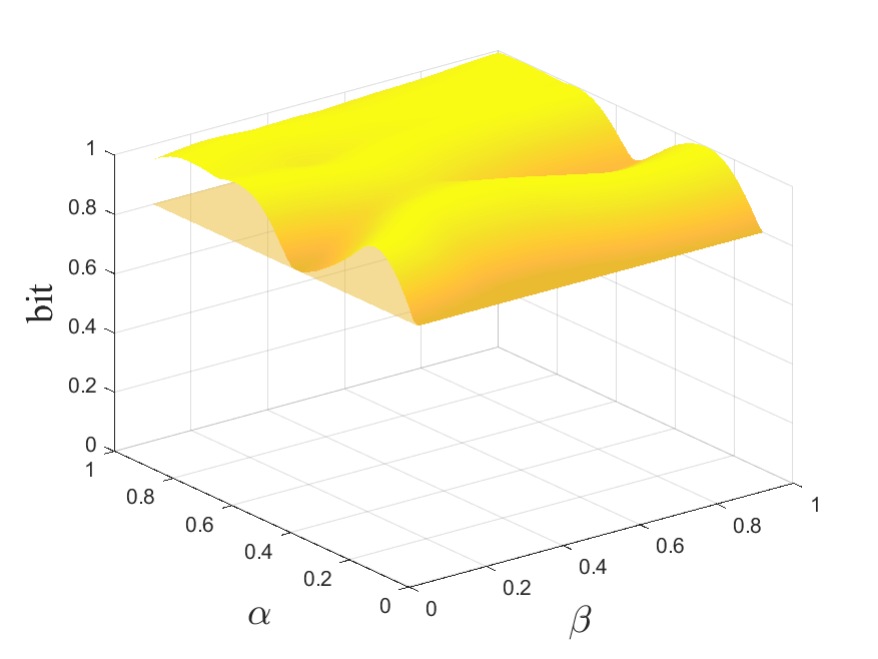}
        }
        \subcaptionbox{$I_0=0.6$}{
            \includegraphics[width=5.1cm]{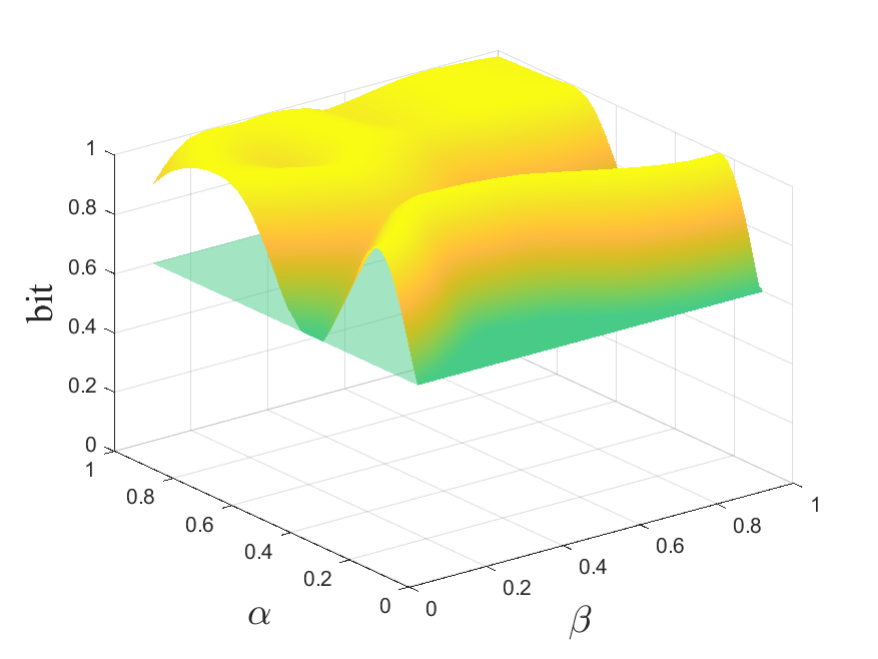}
        }
        \subcaptionbox{$I_0=0.8$}{
            \includegraphics[width=5.1cm]{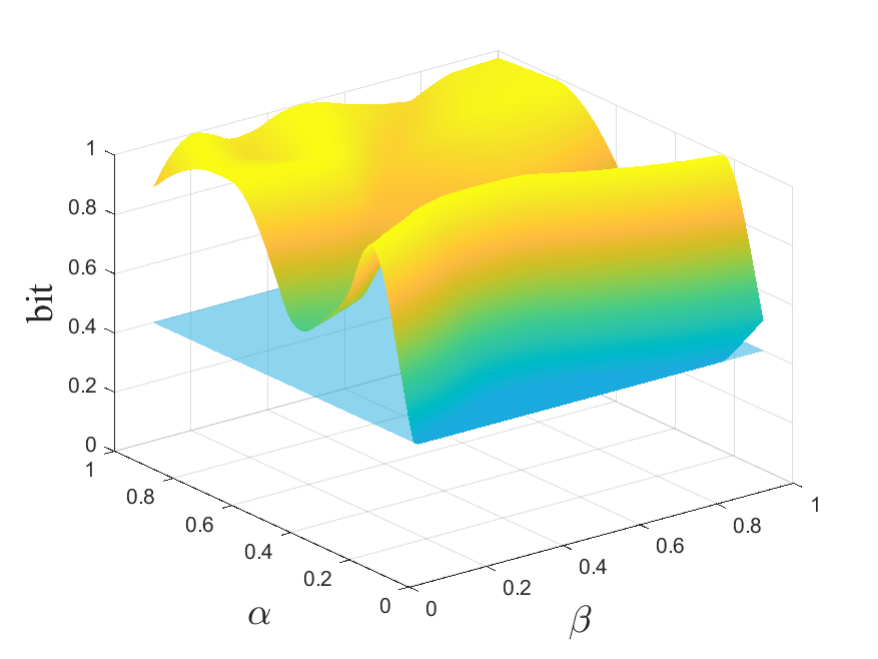}
        }
        \caption{Continuous privacy filters using neural network-based randomizer for HDVs with different privacy levels $I_0$. The x-axis and y-axis represent different sensitive parameters, while the z-axis depicts the entropy of the output distribution corresponding to the respective sensitive parameters. (a), (b) and (c) demonstrate the results with aggregated-level privacy preservation constraints. (d), (e) and (f) are the results with individual-level privacy preservation constraints.}
        \label{fig: continuous privacy filter for HDVs}
\end{figure}

\noindent \emph{Value of individual-level preservation constraints}. To demonstrate the value of considering individual-level preservation constraints, we compare the entropy of the output distribution corresponding to the respective sensitive parameters resulting from two parameter privacy filters: (i) the one with aggregated-level privacy preservation constraints in Eq.~\eqref{eq:opt-randomizer-discrete} and (ii) the one with individual-level preservation constraints in Eq.~\eqref{eq: Individual-Level privacy preservation constraint}. The results are shown in 
Fig. \ref{fig: continuous privacy filter for HDVs}. Comparing Figs. \ref{fig: continuous privacy filter for HDVs} (a)-(c) to Figs. \ref{fig: continuous privacy filter for HDVs} (d)-(f), respectively, we can see that the individual-level privacy preservation constraints in Figs. \ref{fig: continuous privacy filter for HDVs} (d)-(f) enhance the privacy protection of all individual values of the sensitive parameter $\bm{\kappa}$. 

\subsection{Impact on the CAV controller} \label{subsec:control_performance}
In this subsection, we quantify the impact of the privacy filter on the CAV control performance. Specifically, we will first evaluate the magnitude of distortion added to the true states and then evaluate the loss in control performance in terms of fuel consumption and velocity tracking errors.
\subsubsection{Discrepancy Between Pseudo States and True States}
\label{subsubsec: Distribution of the Discrepancy between Pseudo States and True States}
\begin{figure}
    \centering
    \subcaptionbox{}{
        \includegraphics[width = 5.1cm]{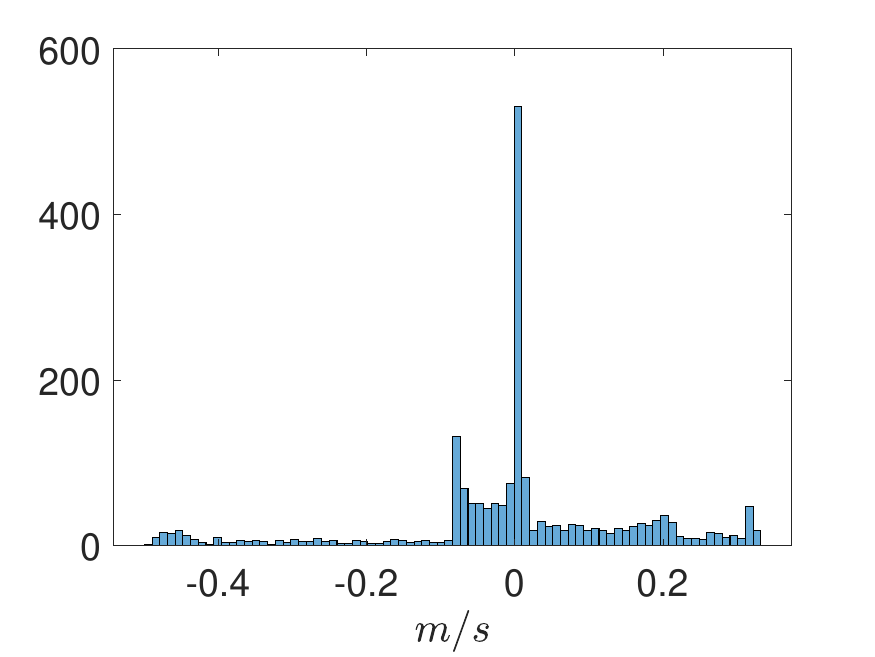}
    }
    \subcaptionbox{}{
        \includegraphics[width = 5.1cm]{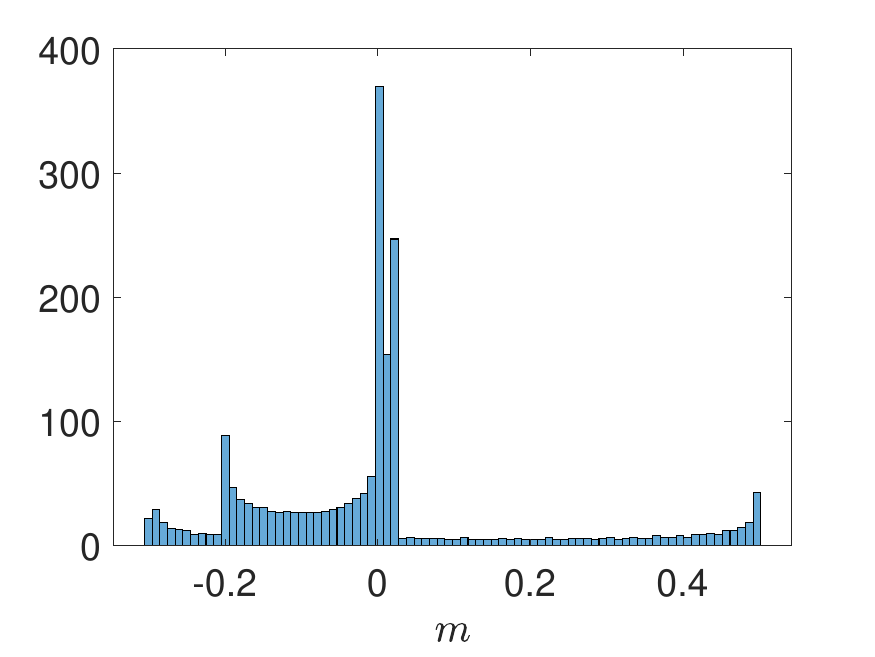}
    }\\
    \subcaptionbox{}{
        \includegraphics[width = 5.1cm]{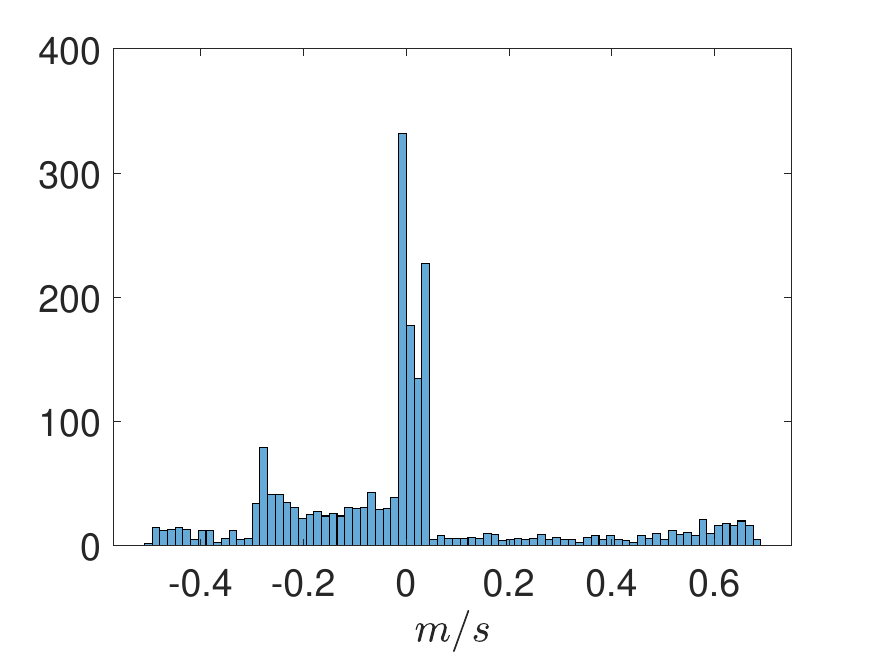}
    }
    \subcaptionbox{}{
        \includegraphics[width = 5.1cm]{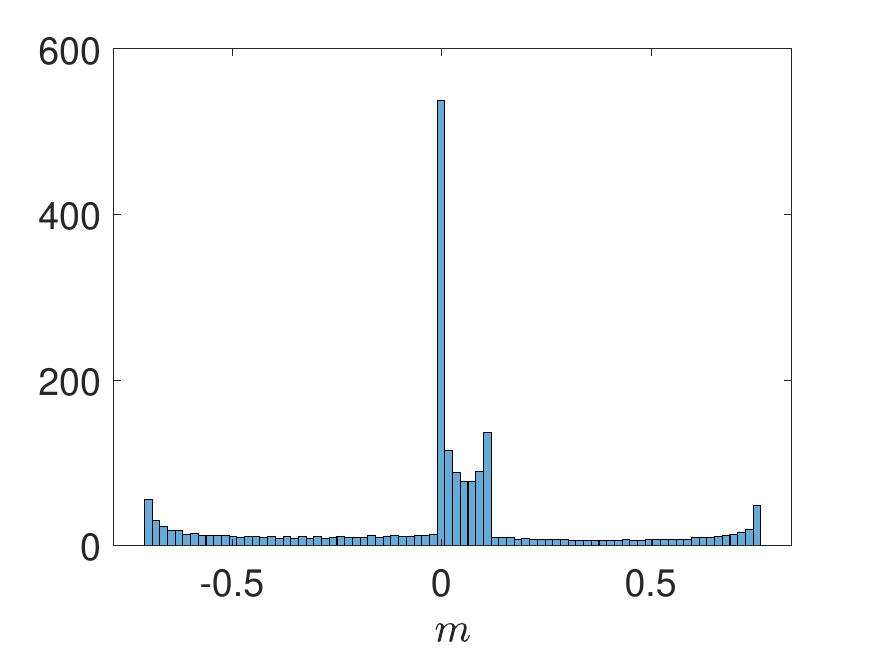}
    }
    \caption{Histogram plots for velocity and spacing discrepancies. (a) and (b) present the velocity and spacing discrepancies in the sine-like velocity disturbance scenario, while (c) and (d) demonstrate the velocity and spacing discrepancies in the emergency braking scenario.}
    \label{fig: Fitted normal distributions for velocity and spacing discrepancies}
\end{figure}

We evaluate the accuracy of the reported states, i.e., the magnitude of discrepancies between the true and the pseudo states, in the two driving scenarios (i.e., sine-like velocity disturbance and emergency braking). 
The simulation extends over a relatively long duration of $1000$ seconds, which, combined with the random acceleration profile, emulates a Monte Carlo method, ensuring that the collected data on discrepancies comprehensively represents a wide range of potential scenarios. The privacy level is chosen as $I_0=0.2$, which is relatively stringent. 
The distributions of the distortions are depicted in Figure \ref{fig: Fitted normal distributions for velocity and spacing discrepancies}, and the means and standard deviations are presented in Table \ref{tab: Fitted normal distributions for velocity and spacing discrepancies}.

The analysis reveals that both velocity and spacing discrepancies have mean values close to $0$ (i.e., almost unbiased) and very low standard deviations, comparable to other sources of noise, such as sensor errors and noises in the communication channels. Such small discrepancies indicate that the impact of the privacy filter on safety and control performance is expected to be low. Moreover, this also reduces the likelihood that the vehicles equipped with the parameter privacy filter are misidentified as attackers. 

\begin{table}[ht]
\caption{Mean and standard deviation of velocity and spacing discrepancies.}
\label{tab: Fitted normal distributions for velocity and spacing discrepancies}
    \centering
    \begin{tabular}{ccccc}
\toprule
& \multicolumn{2}{c}{Sine-Like Velocity Disturbance} & \multicolumn{2}{c}{Emergency Braking}\\\cline{2-5}
    & Mean & Standard deviation & Mean & Standard deviation \\ \midrule
Velocity Discrepancy (m/s) & 8.31e-5   & 0.157 & 1.57e-4   & 0.241\\ 
Spacing Discrepancy (m) &  8.26e-4 &  0.175  &  -6.36e-3    & 0.332\\ 
\bottomrule
\end{tabular}
\end{table}

\subsubsection{Impact on fuel consumption and velocity tracking error} \label{subsubsec:control_performance}

To evaluate the impact of the continuous parameter privacy filter on the CAV controller for mixed-autonomy platoons, we consider two driving scenarios as described in Section \ref{subsec:simu_setting}, each simulated over a period of $30$s. We evaluate the fuel consumption and velocity tracking errors (i.e. the relative velocity difference between the specified vehicle and the head vehicle) to quantify the traffic performance of five conditions: 
\begin{enumerate}
\item[(i)] The scenario without LCC where the CAV adopts the same car-following models as HDVs, which does not require HDVs to share their data.\vspace{-0.6em}
\item[(ii)] The scenario with LCC where the CAV utilizes the DeePC but without the parameter privacy filter,  which requires HDV data but offers no privacy protection.\vspace{-0.6em}
\item[(iii)] The scenario with LCC where the CAV utilizes the DeePC and the parameter privacy filter (applied to all HDVs),  which requires HDV data and offers privacy protection.\vspace{-0.6em} 
\item[(iv)] The scenario with LCC where the CAV utilizes the linear controller but without the parameter privacy filter,  which requires HDV data but offers no privacy protection.\vspace{-0.6em} 
\item[(v)] The scenario with LCC where the CAV utilizes the linear controller and the parameter privacy filter (applied to all HDVs),  which requires HDV data and offers privacy protection. 
\end{enumerate}
The comparison of LCC performance with and without the privacy filter reveals the impact of the parameter privacy filter on the control efficacy of the CAV. Specifically, for the $i$-th vehicle, the fuel consumption rate (mL/s) is calculated based on an instantaneous model proposed in~\citet{bowyer1985guide} given by:
\begin{equation}
f_i= \begin{cases}0.444+0.090 R_i v_i+\left[0.054 a_i^2 v_i\right]_{a_i>0}, & \text { if } R_i>0 \\ 0.444, & \text { if } R_i \leq 0\end{cases}
\end{equation}
with $R_i=0.333+0.00108 v_i^2+1.200 a_i$. The Average Absolute Velocity Error (AAVE) is used to quantify velocity errors, which is obtained by computing the average of $\frac{\left|v_i(t)-v_0(t)\right|}{v_0(t)}$ over all vehicles and the entire simulation period.

\begin{table}[ht]
\centering
\caption{Fuel Consumption and Average Absolute Velocity Error (AAVE)}
\label{tab: Fuel Consumption and Average Absolute Velocity Error (AAVE)}
\begin{tabular}{ccccc}
\toprule
& \multicolumn{2}{c}{Sine-Like Velocity Disturbance} & \multicolumn{2}{c}{Emergency Braking} \\\cline{2-5}
            & Fuel Consumption & AAVE & Fuel Consumption  & AAVE \\
    &{[}mL{]}  && {[}mL{]} &            \\ \midrule
Without LCC   &      690.18               & 0.30  &          251.51       &  0.05\\ 
LCC using DeePC w/o privacy filter &             313.25       & 0.22  &265.82 & 0.20\\ 
LCC using DeePC with privacy filter &               312.99    &  0.22 &      260.96      & 0.20\\ 
LCC using linear controller w/o privacy filter   &          483.07          &  0.27&   238.48   & 0.06\\ 
LCC using linear controller with privacy filter &     496.18               & 0.27 &   239.37   &  0.07  \\
\bottomrule
\end{tabular}
\end{table}
Table~\ref{tab: Fuel Consumption and Average Absolute Velocity Error (AAVE)} presents the fuel consumption and AAVE in different scenarios. Under the sine-like velocity disturbance scenario, it is observed that the adoption of the continuous parameter privacy filter will slightly influence system performance due to the generated noises for both the DeePC and linear feedback controller. However, the resulting system performance is still significantly better than the scenario without LCC, which suggests that the privacy filter does not undermine the benefits of LCC. With the privacy filter applied to LCC using DeePC, we observe a significant reduction in fuel consumption by approximately 43.5\%, but the Average Absolute Velocity Error (AAVE) sees an reduction of about 26.7\%. In the case of the linear controller equipped with the privacy filter, there is also a notable improvement, with fuel consumption dropping by roughly 28.11\% and the AAVE decreasing by around 10\%. Similarly, in the scenario where the head vehicle performs emergency braking, the performance of the controller with privacy filters shows no significant difference compared to the controller without privacy filters.

\section{Head-To-Tail String Stability Analysis}
\label{sec: string stability}
In this section, we perform an analysis of the trade-off between privacy and head-to-tail string stability~\citep{jin2014dynamics} when implementing the parameter privacy filter. 
Specifically, for a sequence of vehicles in a platoon, the definition of head-to-tail string stability is presented below.
\begin{definition}[Head-to-Tail String Stability~\citep{jin2014dynamics}]Given the velocity deviation for the vehicle at the head of a platoon as $v^{\text{err}}_h$ and the one at the tail of a platoon as $v^{\text{err}}_t$, along with their respective Laplace transforms $V^{\text{err}}_h$ and $V^{\text{err}}_t$, the head-to-tail transfer function is defined as follows:
\begin{equation}
    T(s) = \frac{V^{\text{err}}_t}{V^{\text{err}}_h}.
\end{equation}
Let $j^2 = -1$. Then the system is said to be head-to-tail stable if and only if:
\begin{equation}
    |T(j\omega )|^2< 1, \forall \omega \geq 0.
\end{equation} 
\end{definition}

If the platoon has a poor head-to-tail string stability performance~\citep{swaroop1996string}, stop-and-go patterns may emerge in the traffic flow, leading to a significant surge in travel time, fuel consumption, and the potential for accidents. While head-to-tail string stability has been investigated in the LCC~\citep{wang2021leading} to facilitate designing the linear feedback controller, in our scenario, the distortion of the reported states received by the CAV can substantially impact the performance of the CAV controller and hence influence the string stability of the mixed-autonomy platoon system. 

To analyze the influence of different pseudo parameters on head-to-tail string stability, we consider a platoon where only one vehicle is CAV for brevity. The string stability of platoons consisting of more than one CAVs is expected to be further improved since more CAVs will be controllable. We first construct the head-to-tail transfer function for the studied mixed-autonomy platoon in Section~\ref{subsec: TF}, where HDVs are implemented with privacy filters, and the CAV utilizes the linear feedback controller as in Eq.~\eqref{linear controller}. Subsequently, we perform numerical analysis in Section~\ref{subsec: stable_region} to obtain specific head-to-tail string stable regions of different pseudo parameters, systematically exploring the impact of different pseudo parameters on the overall string stability of the system.

\subsection{Head-To-Tail Transfer Function} \label{subsec: TF}
To quantify the impact of the parameter privacy filter on the string stability, we next provide the formulation of the head-to-tail transfer function. 
Similar to \citet{wang2021leading}, we further assume that the HDVs within the platoon are homogeneous. The string stability of platoons with heterogeneous HDVs can be similarly analyzed.  
Let $\Omega_{\mathcal{P}}$ and $\Omega_{\mathcal{F}}$, respectively, represent the sets of the CAV's preceding HDVs and following HDVs, with $|\Omega_{\mathcal{P}}| = n_{\mathcal{P}}$ and $|\Omega_{\mathcal{F}}| = n_{\mathcal{F}}$. The subset of preceding HDVs equipped with privacy filters is denoted by $\tilde{\Omega}_{\mathcal{P}} \subseteq \Omega_{\mathcal{P}}$, and the subset of following HDVs equipped with privacy filters is denoted by $\tilde{\Omega}_{\mathcal{F}} \subseteq \Omega_{\mathcal{F}}$. 
Then, the head-to-tail transfer function of the mixed-autonomy platoon as:
\begin{equation}
    T(s)=G(s)K^{n_{\mathcal{P}}+n_{\mathcal{F}}}(s),
\end{equation}
where $K(s)$ is the local transfer function for the HDV, $G(s)$ is the local tranfer function for the CAV. 

To derive the form of $K(s)$ and $G(s)$, we first present the time-domain representation of the linearized system dynamics (see Appendix~\ref{appendix: Linearized Discrete-Time System Dynamics for Mixed-Autonomy Traffic} for details):
\begin{subequations}
    \begin{align}
    \dot{s}^{\text{err}}_{i}(t) &=v^{\text{err}}_{i-1}(t)-v^{\text{err}}_{i}(t), \quad i\in \Omega \label{eq:lsd_1}\\
    \dot{v}^{\text{err}}_{i}(t) &= 
    \left\{
    \begin{array}{llll}
         &u_i(t), &\quad i \in \Omega_\mathcal{C},\\
         &\alpha_{1} s^{\text{err}}_{i}(t)-\alpha_{2} v^{\text{err}}_{i}(t)+\alpha_{3} v^{\text{err}}_{i-1}(t), &\quad i \in \Omega_\mathcal{H},
    \end{array}
    \right. \label{eq:lsd_2} \\
    \dot{\bar{s}}^{\text{err}}_{i}(t) &=v^{\text{err}}_{i-1}(t)-\bar{v}^{\text{err}}_{i}(t), \quad i\in \tilde{\Omega}_{\mathcal{F}} \cup \tilde{\Omega}_{\mathcal{P}} \label{eq:lsd_3}\\
     \dot{\bar{v}}^{\text{err}}_{i}(t) &=\bar{\alpha}_{1} \bar{s}^{\text{err}}_{i}(t)-\bar{\alpha}_{2} \bar{v}^{\text{err}}_{i}(t)+\bar{\alpha}_{3} v^{\text{err}}_{i-1}(t), \quad i\in \tilde{\Omega}_{\mathcal{F}} \cup \tilde{\Omega}_{\mathcal{P}} \label{eq:lsd_4}
\end{align}\label{eq:linearized system dynamics}
\end{subequations}
where Eqs.\,\eqref{eq:lsd_1}-\eqref{eq:lsd_2} represent the linearized system dynamics, with $\alpha_1, \alpha_2$, and $\alpha_3$ representing the coefficients of the linearized car-following model (see Appendex~\ref{appendix: Linearized Discrete-Time System Dynamics for Mixed-Autonomy Traffic}) as a function of sensitive parameters $\kappa_i$. Eqs.\,\eqref{eq:lsd_3}-\eqref{eq:lsd_4} represent the relation between the pseudo states and the true states, where $\bar{\alpha}_1, \bar{\alpha}_2$, and $\bar{\alpha}_3$ represent the coefficients as a function of the pseudo parameters, and $\bar{s}^{\text{err}}_{i}(t)$ and $\bar{v}^{\text{err}}_{i}(t)$ represent the pseudo states generated by the privacy filter. 

We can perform Laplace transform of Eq.~\eqref{eq:linearized system dynamics} and obtain the local velocity transfer functions for HDV~$i$: 
\begin{subequations}
    \begin{align}
K(s) = \frac{V^{\text{err}}_i}{V^{\text{err}}_{i-1}}=\frac{\alpha_3 s+\alpha_1}{s^2+\alpha_2 s+\alpha_1}, \label{eq:local_TF_HDV} \\
\bar{K}(s)=\frac{\bar{V}^{\text{err}}_i}{V^{\text{err}}_{i-1}} = \frac{\bar{\alpha}_3 s+\bar{\alpha}_1}{s^2+\bar{\alpha}_2 s + \bar{\alpha}_1}. \label{eq:local_TF_HDV_filter}
\end{align}
\end{subequations}
where $V^{\text{err}}_{i}$ and $\bar{V}^{\text{err}}_{i}$ represent
for the true and pseudo velocity deviation of HDV $i$, respectively, and $V^{\text{err}}_{i-1}$ represents 
the Laplace transform for the true velocity deviation of vehicle $i-1$, which serves as input to the dynamic model of HDV $i$.

Recall that the CAV (index $i_{\text{cav}}$) utilizes the linear feedback controller as in Eq.\,\eqref{linear controller}, which can be represented by  
\begin{equation}
\dot v_{i_\text{cav}}(t)=\mu_{i_\text{cav}} s^{\text{err}}_{i_\text{cav}}(t)+k_{i_\text{cav}} v^{\text{err}}_{i_\text{cav}}(t)+\sum_{i \in (\Omega_\mathcal{P} \setminus \tilde{\Omega}_\mathcal{P}) \cup (\Omega_\mathcal{F}\setminus \tilde{\Omega}_\mathcal{F})}\left(\mu_i s^{\text{err}}_i(t)+k_i v^{\text{err}}_i(t)\right)+
\sum_{j \in \tilde{\Omega}_\mathcal{P} \cup \tilde{\Omega}_\mathcal{F}}\left(\mu_j \bar{s}^{\text{err}}_j(t)+k_j \bar{v}^{\text{err}}_j(t)\right),
\end{equation}
where the first two terms represent the feedback to the CAV's own state,  the second term represents the feedback using the true states of HDVs without a privacy filter, and the third term represents the feedback using the pseudo states of HDVs equipped with a privacy filter. 

The details for calculating the transfer function $G(s)$ for the CAV are given in Appendix~\ref{appendix: Derivation of Head-to-Tail Transfer Function}. Here, for presentation simplicity, we provide the result as:
\begin{equation}
    G(s) = \frac{W_2(s)+W_4(s)}{R(s)+W_1(s)+W_3(s)},
\end{equation}
where
\begin{align*}
    &R(s) = \frac{s^2-k_{i_\text{cav}}s-\mu_{i_\text{cav}}}{s},\\
    &W_1(s)=\sum_{i \in \Omega_\mathcal{P} \setminus\tilde{\Omega}_\mathcal{P}}\left(\mu_i\frac{1-K(s)}{sK^{i_\text{cav}-i}(s)}+k_i \frac{1}{K^{i_\text{cav}-i-1}(s)}\right),\\
    &W_2(s)=\sum_{i \in \Omega_\mathcal{F} \setminus\tilde{\Omega}_\mathcal{F}}\left(\mu_i\frac{1}{s}(1-K(s))K^{i-i_\text{cav}-1}(s)+k_i K^{i-i_{\text{cav}}}(s)\right),\\
    &W_3(s)=\sum_{j \in \tilde{\Omega}_{\mathcal{P}}}\left(\mu_j \frac{1-\bar{K}(s)}{sK^{i_\text{cav}-j}(s)}+k_j \frac{\bar{K}(s)}{K^{i_\text{cav}-j(s)}}\right),\\
    &W_4(s)=\sum_{j \in \tilde{\Omega}_{F}}\left(\mu_j \frac{1}{s}\left(1-\bar{K} (s)\right)K^{j-i_{\text{cav}}-1}(s)+k_j \bar{K}(s)K^{j-i_{\text{cav}}-1}(s)\right),
\end{align*} 

Moreover, for two special cases where only one preceding HDV or following HDV is equipped with a privacy filter, we have $W_4 = 0$ or $W_3 = 0$. The head-to-tail transfer function formulation of the two special cases can be given as:
\begin{align}
    T_1(s) &= \frac{W_2(s)}{R(s)+W_1(s)+W_3(s)}K^{n_{\mathcal{P}}+n_{\mathcal{F}}}(s),\label{eq: string stability scenario 1}\\
    T_2(s) &= \frac{W_2(s)+W_4(s)}{R(s)+W_1(s)}K^{n_{\mathcal{P}}+n_{\mathcal{F}}}(s).\label{eq: string stability scenario 2}
\end{align}

For simplicity, we will then analyze the string stable region for the two special cases mentioned above, whereby the string stable condition is met if $|T(j\omega)|\leq 1$.


\subsection{Head-To-Tail String Stable Region}\label{subsec: stable_region}
\begin{figure}[ht]
    \centering
    \subcaptionbox{}{\includegraphics[width=5.5cm]{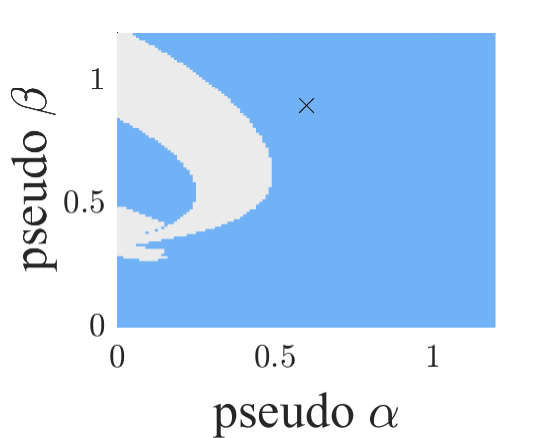}}
    \subcaptionbox{}{\includegraphics[width=5.5cm]{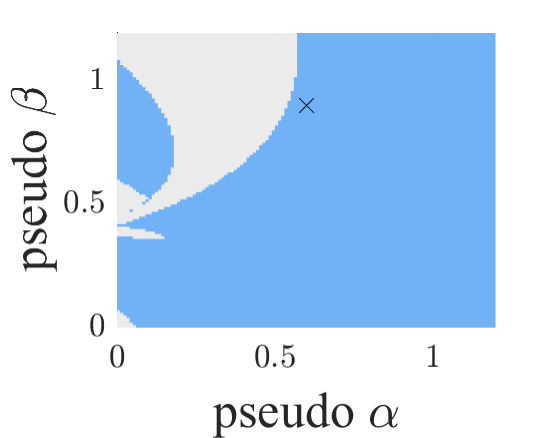}}\\
    \subcaptionbox{}{\includegraphics[width=5.5cm]{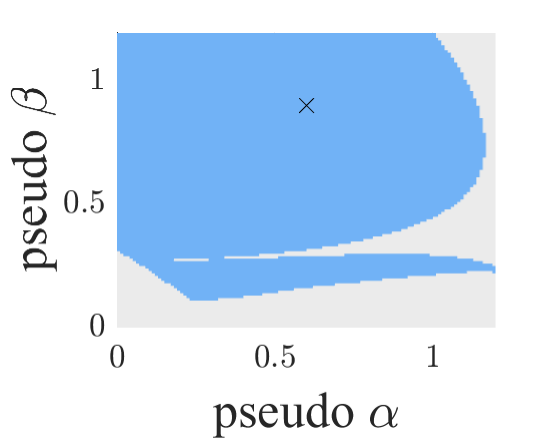}}
    \subcaptionbox{}{\includegraphics[width=5.5cm]{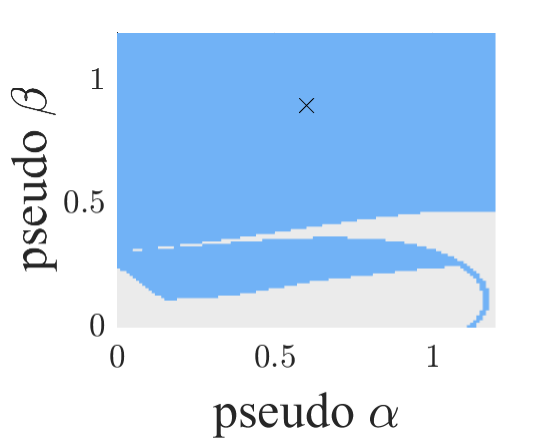}}
    
    \caption{Head-to-tail string stable region within a platoon of $5$ vehicles, with the $3$rd vehicle being a CAV and the remaining vehicles being HDVs, we explore four distinct scenarios. In each case, denoted as (a), (b), (c), and (d), we investigate the string stable region for the $1$-st, $2$-nd, $4$-th, and $5$-th vehicles, respectively, each equipped with a privacy filter. The depicted blue region denotes the string stable area, while the grey region represents the string unstable area. The corresponding parameters for the position with ``$\times$'' indicate the true parameters for the car-following dynamics. The $x$-axis and $y$-axis correspond to the pseudo-sensitive parameters $\alpha$ and $\beta$, both ranging from $0$ to $1.2$.}
    \label{fig: string stable region}
\end{figure}

For simplicity of analysis, we consider a less complex scenario where there is a mixed-autonomy platoon with five vehicles $\Omega=\{1,2,3,4,5\}$, with vehicle 3 being a CAV and the remaining vehicles being HDVs. We perform numerical simulations for four cases, each with an HDV equipped with a privacy filter. 
The results on the head-to-tail string stable regions, as defined by Eq.~\eqref{eq: string stability scenario 1} and Eq.~\eqref{eq: string stability scenario 2}, are illustrated in Fig. \ref{fig: string stable region} for these four cases. 
In this analysis, we maintain constant feedback gains and true sensitive parameters to observe variations in the string stable region under different pseudo parameters. The blue-colored regions in Fig.~\ref{fig: string stable region} denote the string stable regions with pseudo parameters, while the gray-colored region indicates the string unstable region attributed to the influence of pseudo parameters.

To establish a baseline, we select feedback gains that ensure traffic flow stability without a privacy filter. Specifically, the feedback gains for the CAV are set as $\boldsymbol{\mu} = \left[0, -0.15, 0.2, -0.15, -0.1\right]$, and $\boldsymbol{\eta} = \left[0.25, 0.35, -0.5, 0.35, 0.25\right]$. Fig. \ref{fig: string stable region} (a) and (b) depict scenarios where the first HDV and the second HDV are equipped with privacy filters, respectively. Notably, the shapes of the regions are similar, but the unstable region in (b) is larger, attributed to the greater amplitude of feedback gains for the second vehicle compared to the first.

Similarly, Fig. \ref{fig: string stable region} (c) and (d) showcase scenarios where HDV $4$ and HDV $5$ are equipped with privacy filters, respectively. The shapes of the unstable regions are similar, while the unstable region in (d) is larger than in (c) due to the higher amplitude of feedback gains for the HDV $4$'s state compared to HDV $5$.

In summary, the implementation of the devised parameter privacy filter may make originally string-stable 
scenarios string unstable. With fixed control gains, the shape and area of the string unstable regions are influenced by (i) whether the HDV with a privacy filter is following or leading the CAV and (ii) the corresponding amplitude of the state feedback gain of the HDV. Notably, in situations where multiple vehicles are positioned between the HDV with a privacy filter and the CAV, the amplitude of feedback gains for the HDV is typically small in controller design. As a result, the influence on the string unstable region remains limited and can be negligible. Additionally, by judiciously choosing particular pseudo parameters for the privacy filter, it is possible to preserve the head-to-tail string stability within the platoon.

\section{Conclusions and Future Studies} 
\label{sec: Conclusions and Future Studies}
This paper presents a parameter privacy filter to protect the car-following parameters of HDVs in mixed-autonomy platoon control so that adversaries cannot infer these parameters from vehicle states shared by HDVs. The proposed parameter privacy filter distorts the shared vehicle states through two components, i.e., a randomizer and a nonlinear transformation module, whereby the randomizer generates pseudo parameters from the true parameters, and the nonlinear transformation module disguises the HDV's true parameters as the pseudo parameters. In addition to the canonical version that only handles discrete vehicle states, we better tailor the parameter privacy filter to mixed-autonomy platoon control with continuous parameter space by enhancing the computation efficiency via a neural network-based randomizer. Moreover, we further strengthen privacy preservation constraints to provide privacy protection to drivers with each value of the parameter in order to enhance the reliability and practicality of the approach.  Subsequently, we analyze the head-to-tail string stability of the mixed-autonomy platoon when subjected to perturbations from privacy filters. This analysis reveals that the impact of the proposed parameter privacy filter on the string stable region is acceptable. 
The simulation results illustrate that our approach effectively preserves parameter privacy for HDVs with marginal impact on fuel consumption and velocity deviation error. Furthermore, our results show that applying the privacy filter to the preceding vehicle can also help protect the parameter privacy of the following vehicle. Ultimately, by addressing the privacy concerns of HDVs, the proposed methodology has the potential to incentive HDVs to participate in LCC systems and report their data, thereby increasing the amount of data available to the central unit and improving the system performance.  

This research opens several interesting directions for future work. First, we are interested in designing a parameter privacy filter for other transportation systems with a higher dimension of parameters, such as ride-sharing systems. Second, we will extend the privacy filter to protect parameters in networked systems with arbitrary topology, such as scenarios with lane changes. Third, we would like to further enhance the robustness of the learning-based randomizer, e.g., to improve the smoothness of the neural network output by integrating Lipschitz constraints \citep[e.g., LipsNet proposed by][]{song2023lipsnet}.

\printcredits
\section*{Authorship Contribution Statement}
\textbf{Jingyuan Zhou}: Methodology, Validation, Writing - original draft, revised draft. \textbf{Kaidi Yang}: Conceptualization, Methodology, Writing - original draft, revised draft.

\section*{Acknowledgment}
The authors would like to thank Qiqing Wang, Chaopeng Tan, Longhao Yan, and other colleagues in the ITVS lab for fruitful discussions. This study was supported by the Singapore Ministry of Education (MOE) under its Academic Research Fund Tier 1 (A-8000404-01-00). This article solely reflects the opinions and conclusions of its authors and not Singapore MOE or any other entity.

\bibliographystyle{cas-model2-names}

\bibliography{cas-refs}

\appendix
\section{Linearized Discrete-Time System Dynamics for Mixed-Autonomy Traffic}
\label{appendix: Linearized Discrete-Time System Dynamics for Mixed-Autonomy Traffic}
We linearize and discretize the continuous mixed-traffic platoon system presented in Eq.\,\eqref{continuous system} to formulate a discrete-time linear time-invariant (LTI) system characterized by a positive time interval $T_{\text{step}}>0$. Notice that the linearization and time discretization are performed locally on each HV. 

First, we linearize the nonlinear car-following model for HDV $i\in\Omega_{\mathcal{H}}$ in Eq.\,\eqref{eq:system dynamics} using the first-order Taylor expansion around the equilibrium state $x_i^\star = [s_i^\star, v^\star]$, which satisfies $\mathbf{F} _{\mathbb{\kappa}_i}\left(s_i^{\star}, v^\star, v^{\star}\right)=0$ with $v^\star$ specified by the head vehicle. This process yields the following linearized dynamics:
\begin{equation}
\left\{
\begin{aligned}
    &\dot{s}^{\text{err}}_{i}(t) =v^{\text{err}}_{i-1}(t)-v^{\text{err}}_{i}(t), \\
    &\dot{v}^{\text{err}}_{i}(t) =\alpha_{1} s^{\text{err}}_{i}(t)-\alpha_{2} v^{\text{err}}_{i}(t)+\alpha_{3} v^{\text{err}}_{i-1}(t),
\end{aligned}
\right.\label{eq: hdv_linearized}
\end{equation}
where $\alpha_{1}=\frac{\partial F}{\partial s}, \alpha_{2}=\frac{\partial F}{\partial \dot{s}}-\frac{\partial F}{\partial v}, \alpha_{3}=\frac{\partial F}{\partial \dot{s}}$, $s^{\text{err}}_i = s_i - s^\star, v^{\text{err}}_i = v_i - v^\star$ are the error states of the system. The state vector for vehicle $i$ is denoted by $x^{\text{err}}_i(t) = \left[s^{\text{err}}_i(t), v^{\text{err}}_i(t)\right]^\top$. 

Then using the linear dynamics of CAV in Eq.~\eqref{eq:system dynamics} and the linearized HDV dynamics in Eq.~\eqref{eq: hdv_linearized}, the nonlinear system dynamics in Eq.~\eqref{continuous system} can be transformed into the linear form:
\begin{equation}
    \dot{x}^{\text{err}}(t)=A x^{\text{err}}(t)+B u(t)+H \epsilon(t),
    \label{eq: linearized continuous system}
\end{equation}
where $A\in\mathbb{R}^{2n \times 2n}$, $B,H \in \mathbb{R}^{2n \times m}$. $\epsilon(t)$ represents the velocity deviation of the head vehicle, denoted by $v^{\text{err}}_{0}(t)$, which is treated as an exogenous disturbance in the system.

We next transform the continuous-time LTI system represented in Eq.~\eqref{eq: linearized continuous system} to a discrete-time LTI system formulated as:
\begin{equation}
\left\{\begin{array}{l}
x^{\text{err}}(t_s+1)=A_{\mathrm{d}} x^{\text{err}}(t_{\text{s}})+B_{\mathrm{d}} u(t_{\text{s}})+H_{\mathrm{d}} \epsilon(t_{\text{s}}), \\
y(t_{\text{s}})=C_{\mathrm{d}} x^{\text{err}}(t_{\text{s}}),
\end{array}\right.
\label{discrete-time LTI system}
\end{equation}
where  $A_{\mathrm{d}}=e^{A T_{\text{step}}} \in \mathbb{R}^{2 n \times 2 n}$ represents the state transition matrix of the discrete-time system, $B_{\mathrm{d}}=\int_0^{T_{\text{step}}} e^{A t} B d t \in  \mathbb{R}^{2 n \times m}$ is the control input matrix, $ H_{\mathrm{d}}=\int_0^{T_{\text{step}}} e^{A t} H d t \in \mathbb{R}^{2 n}$ corresponds to the disturbance input matrix, and  $C_{\mathrm{d}}=C \in \mathbb{R}^{(n+m) \times 2 n}$ is the output matrix. These matrices collectively define the dynamics of the discrete-time LTI system. 

Additionally, in scenarios where each HDV within the mixed traffic is equipped with a privacy filter as described in Eq.~\eqref{eq: states after privacy filter}, the dynamics of the system are modified to:
\begin{equation}
\left\{\begin{array}{l}
x^{\text{err}}(k+1)=A_{\mathrm{d}} x^{\text{err}}(t_{\text{s}})+B_{\mathrm{d}} u(t_{\text{s}})+H_{\mathrm{d}} \epsilon(t_{\text{s}}), \\
\bar{x}^{\text{err}}_i(t_{\text{s}}) = \Gamma_i\left(x^{\text{err}}_i(t_{\text{s}}) + x^\star,v^{\text{err}}_{i-1}(t_{\text{s}})+v^\star,\kappa_i\right) - x^\star, i \in \Omega_\mathcal{H},\\
y(t_{\text{s}})=C_{\mathrm{d}} x^\prime(t_{\text{s}}),
\end{array}\right.
\label{discrete-time LTI system with privacy filter}
\end{equation}
where $x^\prime(t_{\text{s}}) = \left\{\{\bar{x}^{\text{err}}_i(t)\}_{i \in \Omega_\mathcal{H}},\{x^{\text{err}}_j(t)\}_{j \in \Omega_\mathcal{C}}\right\}$ represents a composite state comprising pseudo states from HDVs and true states from CAVs.

\section{DeePC Formulation for Mixed-Autonomy Platoons}
\label{appendix: DeePC Formulation for Mixed-Autonomy Platoons}

DeePC offers a nonparametric approach that circumvents the need for system identification by directly designing control inputs using pre-collected trajectory data, the details of which are outlined in \cite{coulson2019data,wang2022data}. This pre-collected data sequence includes the control input $u^{\text{d}}$, the head vehicle's velocity error $\epsilon^{\text{d}}$, and the output state $y^{\text{d}}$ over a duration of $T_{\text{ini}}+N$, which is segmented into corresponding "past data" (i.e., $U_{\mathrm{p}}, E_{\mathrm{p}}$, and $Y_{\mathrm{p}}$) of length $T_\text{ini}$ and corresponding "future data" (i.e., $U_{\mathrm{f}}, E_{\mathrm{f}}$, and $Y_{\mathrm{f}}$) of length $N$. Using the online collected data (i.e., $u_{\mathrm{ini}}, \epsilon_{\mathrm{ini}}$, and $y_{\mathrm{ini}}$) of length $T_\text{ini}$ and past data, DeePC calculates the non-parametric system dynamics represented by $g$. The future trajectory of the control input $u$, velocity error $\epsilon$, and output state $y$ is determined using the dynamics $g$ and future data matrices $U_{\mathrm{f}}, E_{\mathrm{f}}, Y_{\mathrm{f}}$. The formulation of DeePC at the $t_{\text{c}}$-th time step is given as:
\begin{subequations}
\begin{align}
\min _{g, u, y, \sigma_y} & \sum_{t_{\text{s}}=t_{\text{c}}}^{t_{\text{c}}+N-1}\left(\|y(t_{\text{s}})\|_Q^2+\|u(t_{\text{s}})\|_R^2\right)+\lambda_g\|g\|_2^2+\lambda_y\left\|\sigma_y\right\|_2^2 \\
\text { subject to } & {\left[\begin{array}{l}
U_{\mathrm{p}} \\
E_{\mathrm{p}} \\
Y_{\mathrm{p}} \\
U_{\mathrm{f}} \\
E_{\mathrm{f}} \\
Y_{\mathrm{f}}
\end{array}\right] g=\left[\begin{array}{c}
u_{\text {ini }} \\
\epsilon_{\text {ini }} \\
y_{\text {ini }} \\
u \\
\epsilon \\
y
\end{array}\right]+\left[\begin{array}{c}
0 \\
0 \\
\sigma_y \\
0 \\
0 \\
0
\end{array}\right],}\label{eq: behavior constraint}\\
&\epsilon=\mathbb{O}_N, \label{eq: future head vehicle velocity error constraint}\\
&s_{\min}-s^\star \leq \mathbb{1}_N \otimes\left[\begin{array}{ll}
\mathbb{O}_{m \times n} & \mathbb{1}_m
\end{array}\right] y \leq s_{\max}-s^\star \label{eq: spacing constraint}\\
&a_{\min } \leq u \leq a_{\max } \label{eq: actuation limitation constraint},
\end{align}
\end{subequations}
where the first term in the objective function penalizes the deviation from the equilibrium state weighted by matrix $Q = \operatorname{diag}(w_v\mathbb{1}_n, w_s\mathbb{1}_n)$, with $w_v$ and $w_s$ indicating the penalty coefficients for velocity and spacing errors, respectively, and $\mathbb{1}_n$ represents an $n-$dimensional identity matrix. Similarly, the second term in the objective function seeks to penalize the energy of the control inputs, weighted by matrix $R = \operatorname{diag}(w_u\mathbb{1}_m)$ with coefficient $w_u$. The third term is introduced to limit the ``complexity'' of the data-driven behavior model to prevent overfitting, while the last term ensures the feasibility of the optimization problem by relaxing the constraints. The coefficients $\lambda_g$ and $\lambda_y$ are positive coefficients that tune the importance of these terms. Eq.~\eqref{eq: behavior constraint} represents the nonparametric model of the mixed-autonomy platoon. Eq.~\eqref{eq: future head vehicle velocity error constraint} indicates that the head vehicle aims to preserve a constant equilibrium velocity in the future. Constraint Eq.~\eqref{eq: spacing constraint} provides lower and upper bounds for the future spacing between the CAV and its preceding vehicle, denoted by $s_{\min}$ and $s_{\max}$, respectively. This is to prevent collisions and dampen traffic disruptions. Finally, constraint Eq.~\eqref{eq: actuation limitation constraint} enforces the actuation limits for each CAV, bounded by the minimum and maximum acceleration values $a_{\min}$ and $a_{\max}$.

\section{Derivation of Head-to-Tail Transfer Function of the CAV}
\label{appendix: Derivation of Head-to-Tail Transfer Function}

We give the details for calculating the local transfer function for the CAV. Recall that the vehicle dynamics of the CAV can be written as Eq.~\eqref{eq:system dynamics} can be represented by
\begin{equation}
\dot v_{i_\text{cav}}(t)=\mu_{i_\text{cav}} s^{\text{err}}_{i_\text{cav}}(t)+k_{i_\text{cav}} v^{\text{err}}_{i_\text{cav}}(t)+\sum_{i \in (\Omega_\mathcal{P} \setminus \tilde{\Omega}_\mathcal{P}) \cup (\Omega_\mathcal{F}\setminus \tilde{\Omega}_\mathcal{F})}\left(\mu_i s^{\text{err}}_i(t)+k_i v^{\text{err}}_i(t)\right)+
\sum_{j \in \tilde{\Omega}_\mathcal{P} \cup \tilde{\Omega}_\mathcal{F}}\left(\mu_j \bar{s}^{\text{err}}_j(t)+k_j \bar{v}^{\text{err}}_j(t)\right),
\end{equation}
The Laplace transform for the velocity dynamics is calculated as follows:  
\begin{equation}
\left(s-k_{i_\text{cav}}-\frac{\mu_{i_\text{cav}}}{s}\right)V^{\text{err}}_{i_\text{cav}} = \sum_{i \in \Omega_\mathcal{P} \cup \Omega_\mathcal{F}\setminus\tilde{\Omega}_\mathcal{P} \cup \tilde{\Omega}_\mathcal{F}}\left(\mu_i S^{\text{err}}_i+k_i V^{\text{err}}_i\right)+\sum_{j \in \tilde{\Omega}_\mathcal{P} \cup \tilde{\Omega}_\mathcal{F}}\left(\mu_j \bar{S}^{\text{err}}_j+k_j \bar{V}^{\text{err}}_j\right).
\label{eq: CAV dynamics laplace}
\end{equation}

Let the local velocity transfer function for CAV be represented as $G(s)=\frac{V^{\text{err}}_{i_\text{cav}}}{ V^{\text{err}}_{i_{\text{cav}}-1}}$. Using the local transfer functions for HDV $K(s)$ in Eq.~\eqref{eq:local_TF_HDV} and the local transfer function for CAV $G(s)$, we can represent the Laplace transformation of the true states of the CAV's preceding and following vehicles as 
\begin{subequations}
\begin{align}
&\left\{
\begin{aligned}
    V^{\text{err}}_i &= \frac{1}{G(s)K^{i_\text{cav}-i-1}(s)}V^{\text{err}}_{i_\text{cav}},\\
    S^{\text{err}}_i &=\frac{1-K(s)}{sG(s)K^{i_\text{cav}-i}(s)}V^{\text{err}}_{i_\text{cav}},
\end{aligned}\quad i \in \Omega_{\mathcal{P}}\setminus\tilde{\Omega}_{\mathcal{P}},
\right.
\label{eq: head to preceding vehicle i laplace} \\
&\left\{
\begin{aligned}
    V^{\text{err}}_i &= K^{i-i_{\text{cav}}}(s)V^{\text{err}}_{i_\text{cav}},\\
    S^{\text{err}}_i &= \frac{1}{s}(1-K(s))K^{i-i_\text{cav}-1}(s)V^{\text{err}}_{i_\text{cav}},
\end{aligned}\quad i \in \Omega_{\mathcal{F}}\setminus\tilde{\Omega}_{\mathcal{F}}.
\right.
\label{eq: head to following vehicle i laplace}
\end{align}
\end{subequations}

Using the local transfer function for HDV with pseudo parameters $\bar{K} (s)$ and the local transfer function for CAV $G(s)$, we can calculate the Laplace transform of the pseudo states of the CAV
\begin{subequations}
\begin{align}
    &\left\{
    \begin{aligned}
        \bar{V}^{\text{err}}_j &=\bar{K}(s)V^{\text{err}}_{j-1}= \frac{\bar{K}(s)}{G(s)K^{i_\text{cav}-j}(s)}V^{\text{err}}_{i_\text{cav}},\\
        \bar{S}^{\text{err}}_j &=\frac{1-\bar{K}(s)}{sG(s)K^{i_\text{cav}-j}(s)}V^{\text{err}}_{i_\text{cav}},
    \end{aligned}\quad j \in \bar{\Omega}_{\mathcal{P}},
    \label{eq: head to preceding vehicle j laplace}
    \right. \\
    &\left\{
    \begin{aligned}
        \bar{V}^{\text{err}}_j &= \bar{K}(s)K^{j-i_{\text{cav}}-1}(s)V^{\text{err}}_{i_\text{cav}},\\
        \bar{S}^{\text{err}}_j &=\frac{1}{s}\left(1-\bar{K} (s)\right)K^{j-i_{\text{cav}}-1}(s)V^{\text{err}}_{i_\text{cav}},
    \end{aligned}\quad j \in \bar{\Omega}_{\mathcal{F}}.
    \right.
    \label{eq: head to following vehicle j laplace}
\end{align}
\end{subequations}

Then integrating Eq.~\eqref{eq: head to following vehicle j laplace} and Eq.~\eqref{eq: head to preceding vehicle j laplace} into Eq.~\eqref{eq: CAV dynamics laplace}, we have:
\begin{equation}
    G(s) = \frac{W_2(s)+W_4(s)}{R(s)+W_1(s)+W_3(s)}.
\end{equation}
where
\begin{align*}
    &R(s) = \frac{s^2-k_{i_\text{cav}}s-\mu_{i_\text{cav}}}{s},\\
    &W_1(s)=\sum_{i \in \Omega_\mathcal{P} \setminus\tilde{\Omega}_\mathcal{P}}\left(\mu_i\frac{1-K(s)}{sK^{i_\text{cav}-i}(s)}+k_i \frac{1}{K^{i_\text{cav}-i-1}(s)}\right),\\
    &W_2(s)=\sum_{i \in \Omega_\mathcal{F} \setminus\tilde{\Omega}_\mathcal{F}}\left(\mu_i\frac{1}{s}(1-K(s))K^{i-i_\text{cav}-1}(s)+k_i K^{i-i_{\text{cav}}}(s)\right),\\
    &W_3(s)=\sum_{j \in \tilde{\Omega}_{\mathcal{P}}}\left(\mu_j \frac{1-\bar{K}(s)}{sK^{i_\text{cav}-j}(s)}+k_j \frac{\bar{K}(s)}{K^{i_\text{cav}-j(s)}}\right),\\
    &W_4(s)=\sum_{j \in \tilde{\Omega}_{F}}\left(\mu_j \frac{1}{s}\left(1-\bar{K} (s)\right)K^{j-i_{\text{cav}}-1}(s)+k_j \bar{K}(s)K^{j-i_{\text{cav}}-1}(s)\right).\\
\end{align*}

\end{document}